\documentclass{article}
\begin{document}
{\Large
\begin{center}
{\bf
Excited ${\bf (70,L^+)}$ baryon resonances in the relativistic quark
model}
\end{center}
}
\vskip3ex
\noindent
S.M. Gerasyuta and E.E. Matskevich

\vskip2ex
\noindent
Department of Theoretical Physics, St. Petersburg State University,
198904, St. Petersburg, Russia

\noindent
Department of Physics, LTA, 194021, St. Petersburg, Russia

\vskip4ex
\begin{center}
{\bf Abstract}
\end{center}
\vskip4ex
The masses of positive parity $(70,0^+)$ and $(70,2^+)$ nonstrange and
strange baryons are calculated in the relativistic quark model.
The relativistic three-quark equations of the $(70,L^+)$ multiplets
are found in the framework of the dispersion relation technique.
The approximate solutions of these equations using the method based
on the extraction of leading singularities of the amplitude are
obtained. The calculated mass values of the $(70,L^+)$ multiplets
are in good agreement with the experimental ones.
\vskip2ex
\noindent
PACS: 11.55.Fv, 12.39.Ki, 12.40.Yx, 14.20.-c.
\vskip2ex
\noindent
e-mail address: gerasyuta@SG6488.spb.edu

\noindent
e-mail address: matskev@pobox.spbu.ru
\vskip2ex
{\large \bf 1. Introduction.}
\vskip2ex

Hadron spectroscopy has always played an important role in the revealing
mechanisms underlying the dynamic of strong interactions.

At low energies, typical for baryon spectroscopy, QCD does not admit
a perturbative expansion in the strong coupling constant. In 1974
't Hooft [1] suggested a perturbative expansion of QCD in terms of the
parameter $1/N_c$ where $N_c$ is the number of colors. This suggestion
together with the power counting rules of Witten [2] has lead to the
$1/N_c$ expansion method which allows to systematically analyse baryon
properties. The success of the method stems from the discovery that
the ground state baryons have an exact contracted $SU(2N_f)$ symmetry
when $N_c \to \infty$ [3, 4], $N_f$ being the number of flavors. For
$N_c \to \infty$ the baryon masses are degenerated. For large $N_c$ the
mass splitting starts at order $1/N_c$. Operator reduction rules simplify
the $1/N_c$ expansion [5, 6].

A considerable amount of work has been devoted to the ground state
baryons, described by the symmetric representation $56$ of $SU(6)$
[7-11]. The excited baryons belonging to $(56,L)$ multiplets can be
studied by analogy with the ground state. In this case both the orbital
and the spin-flavor parts of the wave functions are symmetric. Explicit
forms for such wave functions were given, for example, in Ref [12].
Together with color part, they generate antisymmetric wave functions.
Naturally, it turned out that the splitting starts at order $1/N_c$
as for the ground state.

The states belonging to $(70,L)$ multiplets are apparently more
difficult. In this case the general practice was to split the baryon
into excited quark and a symmetric core, the latter being either in
the ground state for the $N=1$ or in an excited state for the
$N\ge 2$ bands. Recently Matagne and Stancu have suggested the new
approach [13] for the excited $(70,1^-)$ multiplet. They solved
the problem by removing the splitting of generators and using
orbital-flavor-spin wave functions. The excited baryons are considered as
bound states. The basic conclusion is that the first order correction to
the baryon masses is order $1/N_c$ instead of order $N_c^0$ as previously
found. The conceptual difference between the ground state and the excited
states is therefore removed.

The constituent quark model suggests that $(70,L^+)$ baryons are
composed of the system of $N_c$ quarks, which is divided into an excited
quark and a core, which can be excited or not [14]. The standart procedure,
used in Ref. [15], for calculating the mass spectrum is to reduce the
wave function to that of a product of a symmetric orbital and symmetric
flavor-spin wave function for the core of $N_c-1$ quarks times the wave
function of the excited quark. This implies that the total
orbital-flavor-spin wave function is truncated to a single term,
described by the product of two Young tableaux, each with the excited
quark in the second row. Many others terms, related to Young tableaux
with the excited quark in the first row, are neglected [16].
At the same time each $SU(6)$ and $O(3)$ generators is splitted into
two terms, are acting on the core and the other on the excited quark.
It is not possible to treat the $(70,L^+)$ multiplet composed of
strange baryons without simplifying the baryon wave function and
splitting of $SU(6)$ generators.

In the series of papers [17-21] a practical treatment of relativistic
three-hadron systems have been developed. The physics of three-hadron
system is usefully described in term of the pairwise interactions
among the three particles. The theory is based on the two principles
of unitarity and analyticity, as applied to the two-body subenergy
channels. The linear integral equations in a single variable are obtained
for the isobar amplitudes. Instead of the quadrature methods of obtaining
solution the set of suitable functions is identified and used as basis
set for the expansion of the desired solutions. By this means the couple
integral equations are solved in terms of simple algebra.

In our papers [22, 23] relativistic generalization of the three-body
Faddeev equations was obtained in the form of dispersion relations in the
pair energy of two interacting particles. The mass spectrum of $S$-wave
baryons including $u$, $d$, $s$-quarks was calculated by a method based
on isolating the leading singularities in the amplitude. We searched for
the approximate solution of integral three-quark equations by taking
into account two-particle and triangle singularities, all the weaker ones
being neglected. If we considered such an approximation, which corresponds
to taking into account two-body and triangle singularities, and defined
all the smooth functions of the middle point of the physical
region of Dalitz-plot, then the problem was reduced to the one of solving
a system of simple algebraic equations.

In our paper [24] the construction of the orbital-flavor-spin wave functions
for the $(70,1^-)$ multiplet are given. We deal with a three-quark system
having one unit of orbital excitation. The orbital part of wave function
must have a mixed symmetry. The spin-flavor part of wave function must
have the same symmetry is order to obtain a totally symmetric state in the
orbital-flavor-spin space. The integral equations using the
orbital-flavor-spin wave functions was constructed. It allows to calculate
the mass spectra for all baryons of $(70,1^-)$ multiplet. We take into
account the $u$, $d$, $s$-quarks. We have represented the $30$ nonstange and
strange resonances belonging to the $(70,1^-)$ multiplet. The $15$
resonances are in good agreement with experimental data [25]. We have
predicted $15$ masses of baryons. In our model the four parameter
are used: gluon coupling constants $g_{+}$ and $g_{-}$ for the various
parity, cutoff energy parameters $\lambda$, $\lambda_s$ for the
nonstrange and strange diquarks.

The paper is organized as follows. After this introduction, we discuss
the construction of the orbital-flavor-spin wave functions for the
$(70,0^+)$ and $(70,2^+)$ multiplets.

In Sect. 3 the relativistic three-quark equations are obtained in
the form of the dispersion relation over the two-body subenergy.

In Sect. 4 the systems of equations for the reduced amplitudes are
derived.

Section 5 is devoted to the calculation results for the mass spectrum
of the $(70,0^+)$ and $(70,2^+)$ multiplets (Tables I-XII).

In Conclusion, the status of the considered model is discussed.

In Appendix I the wave functions of $(70,0^+)$ and $(70,2^+)$ baryon
resonances are given.

In Appendix II the reduced equations for the $(70,0^+)$ and $(70,2^+)$
multiplets are obtained.

\vskip2ex
{\large \bf 2. The wave function of $(70,0^+)$ and $(70,2^+)$ excited
states.}
\vskip2ex

The multiplet $(70,2^+)$ consists of the excited baryon resonances with
the orbital moment $L=2$ and positive-parity. According to the
nonrelativistic approach [15], the $(70,2^+)$ multiplet states includes
the two quarks on the $1s$-levels and one quark on the $1d$-level ($002$)
or the two quarks on the $1p$-levels and one quark on the $1s$-level
($110$). Then the baryons of multiplet $(70,2^+)$ consists of the
superposition of the $002$ and $110$ states. The transition of these
states with the orbital moment $L_z=2$ are considered.

The multiplet $(70,2^+)$ of $SU(6)$ includes the decuplet $(10,2)$
with the spin $S=\frac{1}{2}$, octet $(8,2)$ with the spin
$S=\frac{1}{2}$, octet $(8,4)$ with the spin $S=\frac{3}{2}$ and singlet
$(1,2)$ with the spin $S=\frac{1}{2}$. Taking into account the orbital
and spin moment $\overrightarrow{J}=\overrightarrow{L}+\overrightarrow{S}$,
we obtain the angular moment for the $S=\frac{1}{2}$
$J=\frac{3}{2},\, \frac{5}{2}$ and for the $S=\frac{3}{2}$
$J=\frac{1}{2},\, \frac{3}{2},\, \frac{5}{2},\, \frac{7}{2}$. We can
represent the total multiplet $(70,2^+)$:

\vskip2.5ex
$(10,2): \hskip7.0ex \frac{3}{2}^+ , \hskip3.0ex \frac{5}{2}^+
\hskip22.0ex P_{33},\hskip3.0ex F_{35}$

\vskip2.5ex
$(8,2): \hskip8.5ex \frac{3}{2}^+ , \hskip3.0ex \frac{5}{2}^+
\hskip22.0ex P_{13},\hskip3.0ex F_{15}$

\vskip2.5ex
$(8,4): \hskip8.5ex \frac{1}{2}^+ , \hskip3.0ex \frac{3}{2}^+ ,
\hskip3.0ex \frac{5}{2}^+ ,\hskip3.0ex \frac{7}{2}^+
\hskip7.6ex P_{11},\hskip3.0ex P_{13},\hskip3.0ex F_{15},\hskip3.0ex F_{17}$

\vskip2.5ex
$(1,2): \hskip8.5ex \frac{3}{2}^+ , \hskip3.0ex \frac{5}{2}^+
\hskip22.0ex P_{03},\hskip3.0ex F_{05}$

\vskip2.5ex
The $(70,2^+)$ multiplet includes the $34$ baryons with different masses.

The $(70,0^+)$ multiplet includes the excited baryon resonances with
the orbital moment $L=0$ and the positive-parity. The states of this
multiplet consist of the two quarks on the $1s$-levels and one radial
excited quark on the level $2s$, or the two quarks on the $1p$-levels
with the projection of orbital moment $L_z=0$ and one quark on the
$1s$-level. We consider the spin $S=\frac{1}{2}$ and $J=\frac{1}{2}$,
and $S=\frac{3}{2}$, $J=\frac{3}{2}$. We can represent the total
multiplet $(70,0^+)$:

\vskip2.5ex

$(10,2): \hskip7.0ex \frac{1}{2}^+ \hskip29.0ex P_{31}$

\vskip2.5ex
$(8,2): \hskip8.5ex \frac{1}{2}^+ \hskip29.0ex P_{11}$

\vskip2.5ex
$(8,4): \hskip8.5ex \frac{3}{2}^+ \hskip29.0ex P_{13}$

\vskip2.5ex
$(1,2): \hskip8.5ex \frac{1}{2}^+ \hskip29.0ex P_{01}$

\vskip2.5ex
The $(70,0^+)$ multiplet includes the $13$ baryons with different masses.

The three-quark wave function of the excited baryon posseses the symmetry
$SU(6)\times O(3)\times SU(3)_{c}$, where the $SU(3)_{c}$ group determines
the color symmetry, therefore the total wave function is antisymmetric.
The part of wave function $SU(6)\times O(3)$ must be total symmetric.

The $O(3)$ wave functions with the mixed symmetry allow to construct two
states with the mixed symmetry and the positive-parity. Then we use these
states and two mixed multiplets $70$ and $70^{'}$ of group $SU(6)$.
We can construct the total symmetric state of multiplet $(70,2^+)$.

$O(3)$ wave functions with the mixed symmetry are there:

$$\varphi_{MA}^{O(3)}=\frac{1}{2}\left(
020-200-101+011
\right),\quad\quad
\varphi_{MS}^{O(3)}=\frac{1}{\sqrt{12}}\left(
020+200-2\cdot 002-101-011+2\cdot 110
\right),\eqno (1)$$

\noindent
here $0, 1, 2$ are the values of the projections of quark orbital momentum.
$MA$ and $MS$ correspond to the mixed antisymmetric and symmetric part
of wave function. $SU(6)$ wave functions are chosen for the each states.
For the sake of simplicity we have derived the wave functions for the
$(10,2)$ decuplet $\Sigma^+$.

By analogy of the paper [24]:

$$\varphi_{MA}^{SU(6)}=\varphi_{S}^{SU(3)}\varphi_{MA}^{SU(2)},\quad\quad
\varphi_{MS}^{SU(6)}=\varphi_{S}^{SU(3)}\varphi_{MS}^{SU(2)},\eqno (2)$$

$$\varphi_{MA}^{SU(2)}=\frac{1}{\sqrt{2}}\left(
\uparrow \downarrow \uparrow-\downarrow \uparrow \uparrow
\right),\quad\quad
\varphi_{MS}^{SU(2)}=\frac{1}{\sqrt{6}}\left(
\uparrow \downarrow \uparrow+\downarrow \uparrow \uparrow-
2\uparrow \uparrow \downarrow
\right),\eqno (3)$$

$$\varphi_{S}^{SU(3)}=\frac{1}{\sqrt{3}}\left(
usu+suu+uus
\right).\eqno (4)$$

$\uparrow$ and $\downarrow$ determine the spin directions. $2$, $1$ and $0$
correspond to the excited or nonexcited quarks.

We use the functions (2) -- (4) and construct the $SU(3)$ functions
for each particle.

The total symmetric $SU(6)\times O(3)$ wave functions are similar to:

$$\varphi=\frac{1}{\sqrt{2}}\left(
\varphi_{MA}^{SU(6)}\varphi_{MA}^{O(3)}+
\varphi_{MS}^{SU(6)}\varphi_{MS}^{O(3)}
\right)\equiv\frac{1}{\sqrt{2}}\varphi_{S}^{SU(3)}\left(
\varphi_{MA}^{SU(2)}\varphi_{MA}^{O(3)}+
\varphi_{MS}^{SU(2)}\varphi_{MS}^{O(3)}
\right).\eqno (5)$$

For instance, we have obtained the wave function for the $\Sigma^+$ of
$(10,2)$ multiplet as:

$$\varphi_{\Sigma^{+}(10,2)}=\frac{\sqrt{3}}{18}\left(
2\{u^2\downarrow u\uparrow s\uparrow\}+
\{s^2\downarrow u\uparrow u\uparrow\}-
\{u^2\uparrow u\downarrow s\uparrow\}-
\{u^2\uparrow u\uparrow s\downarrow \}-
\{s^2\uparrow u\uparrow u\downarrow \}-\right.$$
$$\left.-2\{u\downarrow u^1\uparrow s^1\uparrow\}-
\{s\downarrow u^1\uparrow u^1\uparrow\}+
\{u\uparrow u^1\downarrow s^1\uparrow\}+
\{u\uparrow u^1\uparrow s^1\downarrow \}+
\{s\uparrow u^1\uparrow u^1\downarrow \}
\right).\eqno (6)$$

Here the parenthesises determine the symmetrical function:

$$\{{abc}\}\equiv abc+acb+bac+cab+bca+cba.\eqno (7)$$

The wave functions of $\Sigma^{0}$- ¨ $\Sigma^{-}$-hyperons can be
constructed by similar way.

For the $\Delta$ baryon of $(10,2)$ multiplet the wave function can be
obtained if we replace by $u\leftrightarrow s$ quarks.

$$\varphi_{\Delta^{++}(10,2)}=\frac{1}{6}\left(
\{u^2\downarrow u\uparrow u\uparrow\}-\{u^2\uparrow u\uparrow u\downarrow\}
-\{u\downarrow u^1\uparrow u^1\uparrow\}
+\{u\uparrow u^1\uparrow u^1\downarrow\}
\right).\eqno (8)$$

For the $\Xi^{0,-}$ of the $(10,2)$ multiplet the wave function is
similar to the $\Sigma^{+,-}$ state with the replacement by
$u\leftrightarrow s$ or $d\leftrightarrow s$. The wave function for the
$\Omega^{-}$ of the $(10,2)$ decuplet is determined as the $\Delta^{++}$
state with the replacement by $u\rightarrow s$ quarks.

\newpage
{\large \bf 3. The three-quark integral equations for the $(70,0^+)$
and $(70,2^+)$ multiplets.}
\vskip2.0ex

By consideration of the construction of $(70,0^+)$ and $(70,2^+)$
multiplets integral equations we need to using the projectors for the
different diquark states. The projectors to the symmetric and antisymmetric
states can be obtained as:

$$\frac{1}{2}\left(q_1 q_2+q_2 q_1\right),\quad\quad
\frac{1}{2}\left(q_1 q_2-q_2 q_1\right).\eqno (9)$$

The spin projectors are following:

$$\frac{1}{2}\left(\uparrow\downarrow+\downarrow\uparrow
\right),\quad\quad
\frac{1}{2}\left(\uparrow\downarrow-\downarrow\uparrow
\right).\eqno (10)$$

The orbital moment excitation projectors take into account the transition
of diquarks $20\leftrightarrow 11$ with the value of orbital moment
projection $L_z=2$.

\vskip2.5ex
$L_z=2:$

$$200: \quad  A^{s0+}, \,\,  A^{d2+}, \quad\quad\quad
011: \quad  A^{p2+}, \,\,  A^{p1-}. \hskip40ex\eqno (11)$$

$L_z=1:$

$$1^*00: \quad  A^{s0+}, \,\,  A^{d1+}, \quad\quad\quad
00^*1: \quad  A^{p1+}, \,\,
\frac{1}{2}(A^{p0-}+A^{p1-}).\hskip27ex\eqno (12)$$

$L_z=0:$

$$0^*00: \quad  A^{s0+}, \,\,  A^{d0+}, \quad\quad\quad
00^*0^*,\,\, 01(-1)^*: \quad  A^{p0+}, \,\,
\frac{1}{4}(2\cdot A^{p0-}+A^{p1-}+A^{p(-1)-}).\hskip2ex\eqno (13)$$

$L_z=-1:$

$$(-1)^*00: \quad  A^{s0+}, \,\,  A^{d(-1)+},
\quad\quad\quad
00^*(-1)^*: \quad  A^{p(-1)+}, \,\,
\frac{1}{2}(A^{p0-}+A^{p(-1)-}).\hskip10ex\eqno (14)$$

The upper index determines the diquark states. The excited quark is
determined by (*).

\noindent
$11\hskip1.0ex :$

$$\frac{A^{sym}}{4}
\left(2\cdot 11+20+02\right), \eqno (15)$$

\noindent
$10\hskip1.0ex :$

$$\frac{A^{sym}}{2}
\left(10+01\right)
+\frac{A^{asym}}{2}
\left(10-01\right), \eqno (16)$$

\noindent
$00\hskip1.0ex :$

$$ A^{sym} \cdot 00, \eqno (17)$$

\noindent
$20\hskip1.0ex :$

$$\frac{A^{sym}}{4}
\left(20+02+2\cdot 11\right)
+\frac{A^{asym}}{2}\left(20-02\right), \eqno (18)$$

\noindent
here

$$A^{sym}=\frac{A^{s0+}+A^{p2+}}{2}, \quad
\quad A^{asym}=\frac{A^{d2+}+A^{p1-}}{2}. \eqno (19)$$

The product of three projectors $SU(3)_{f} \times SU(2) \times O(3)$
must be symmetrical.

For example, the projector to the diquark $u^2 \uparrow s\downarrow$
is following:

$$\frac{A^{sym\, s}_1}{16}(us+su)
(\uparrow\downarrow+\downarrow\uparrow)(20+02+2\cdot 11)+
\frac{A^{sym\, s}_0}{16}(us-su)
(\uparrow\downarrow-\downarrow\uparrow)(20+02+2\cdot 11)$$

$$+\frac{A^{asym\, s}_1}{8}(us-su)
(\uparrow\downarrow+\downarrow\uparrow)(20-02)+
\frac{A^{asym\, s}_0}{8}(us+su)
(\uparrow\downarrow-\downarrow\uparrow)(20-02).\eqno(20)$$

\noindent
here the lower index of amplitude corresponds to the diquark spin
($1$ or $0$), and the upper index $s$ points out the strangeness of diquark.

For the sake of simplicity we derive the relativistic Faddeev equations
using the $\Sigma$ hyperon with $J^p=\frac{5}{2} ^{+}$ of the (10,2)
multiplets. We use the graphic equations for the functions $A_J(s,s_{ik})$
[22, 23]. In order to represent the amplitude $A_J(s,s_{ik})$ in the form
of dispersion relation, it is necessary to define the amplitudes of
quark-quark interaction $a_J(s_{ik})$. The pair quarks amplitudes
$qq\rightarrow qq$ are calculated in the framework of the dispersion
$N/D$ method with the input four-fermion interaction with quantum numbers
of the gluon [26]. We use results of our relativistic quark model [27]
and write down the pair quark amplitudes in the form:

$$a_J(s_{ik})=\frac{G^2_J(s'_{ik})}
{1-B_J(s_{ik})},\eqno (21)$$

$$B_J(s_{ik})=\int\limits_{(m_i+m_k)^2}^{\Lambda_J(i,k)}
\frac{ds'_{ik}}{\pi}\frac{\rho_J(s'_{ik})G^2_J(s'_{ik})}
{s'_{ik}-s_{ik}},\eqno (22)$$

$$\rho_J (s_{ik})=\frac{(m_i+m_k)^2}{4\pi}
\left(\alpha_J\frac{s_{ik}}{(m_i+m_k)^2}
+\beta_J +\frac{\delta_J}{s_{ik}} \right)\times$$

$$\times\frac{\sqrt{(s_{ik}-(m_i+m_k)^2)(s_{ik}-(m_i-m_k)^2)}}
{s_{ik}}\, .\eqno (23)$$

Here $G_J$ is the diquark vertex function; $B_J(s_{ik})$, $\rho_J (s_{ik})$
are the Chew-Mandelstam function [28] and the phase space consequently.
$s_{ik}$ is the two-particle subenergy squared (i,k=1,2,3), $s$ is the
systems total energy squared. $\Lambda_J(i,k)$ is the pair energy cutoff.
The coefficient of Chew-Mandelstam function are given in Table XIII.

In the case in question the interacting quarks do not produce bound state,
then the integration in dispersion integrals is carried out from
$(m_i+m_k)^2$ to $\Lambda_J(i,k)$. All diagrams are classified over
the last quark pair (Fig.1).

\vskip60pt
\begin{picture}(600,60)
\put(-10,40){\line(1,0){33}}
\put(-10,50){\line(1,0){28}}
\put(-10,60){\line(1,0){33}}
\put(19,46){\line(1,1){15}}
\put(22,41){\line(1,1){17}}
\put(27.5,38.5){\line(1,1){14}}
\put(41,56){\vector(2,1){28}}
\put(42.5,50){\vector(1,0){35}}
\put(41,44){\vector(2,-1){28}}
\put(30,50){\circle{25}}
\put(70,78){$1$}
\put(70,55){$2$}
\put(70,20){$3$}
\put(87,47){$=$}
\put(107,53){\line(1,0){28}}
\put(107,50){\line(1,0){28}}
\put(107,47){\line(1,0){28}}
\put(135,53){\vector(2,1){28}}
\put(135,50){\vector(1,0){35}}
\put(135,47){\vector(2,-1){28}}
\put(163,78){$1$}
\put(163,55){$2$}
\put(163,20){$3$}
\put(180,47){$+$}
\put(200,40){\line(1,0){33}}
\put(200,50){\line(1,0){28}}
\put(200,60){\line(1,0){33}}
\put(229,46){\line(1,1){15}}
\put(232,41){\line(1,1){17}}
\put(237.5,38.5){\line(1,1){14}}
\put(251,44){\vector(2,-1){28}}
\put(240,50){\circle{25}}
\put(268,54){\oval(33,33)[tl]}
\put(252,70){\oval(33,33)[br]}
\put(269,71){\vector(2,3){15}}
\put(269,71){\vector(2,-1){24}}
\put(287,95){$1$}
\put(295,65){$2$}
\put(280,20){$3$}
\put(300,47){$+$}
\end{picture}

\vskip60pt
\begin{picture}(600,60)
\put(90,47){$+$}
\put(110,40){\line(1,0){33}}
\put(110,50){\line(1,0){28}}
\put(110,60){\line(1,0){33}}
\put(139,46){\line(1,1){15}}
\put(142,41){\line(1,1){17}}
\put(147.5,38.5){\line(1,1){14}}
\put(161,44){\vector(2,-1){28}}
\put(150,50){\circle{25}}
\put(178,54){\oval(33,33)[tl]}
\put(162,70){\oval(33,33)[br]}
\put(179,71){\vector(2,3){15}}
\put(179,71){\vector(2,-1){24}}
\put(197,95){$1$}
\put(205,65){$3$}
\put(190,20){$2$}
\put(210,47){$+$}
\put(230,40){\line(1,0){33}}
\put(230,50){\line(1,0){28}}
\put(230,60){\line(1,0){33}}
\put(259,46){\line(1,1){15}}
\put(262,41){\line(1,1){17}}
\put(267.5,38.5){\line(1,1){14}}
\put(281,44){\vector(2,-1){28}}
\put(270,50){\circle{25}}
\put(298,54){\oval(33,33)[tl]}
\put(282,70){\oval(33,33)[br]}
\put(299,71){\vector(2,3){15}}
\put(299,71){\vector(2,-1){24}}
\put(317,95){$2$}
\put(325,65){$3$}
\put(310,20){$1$}
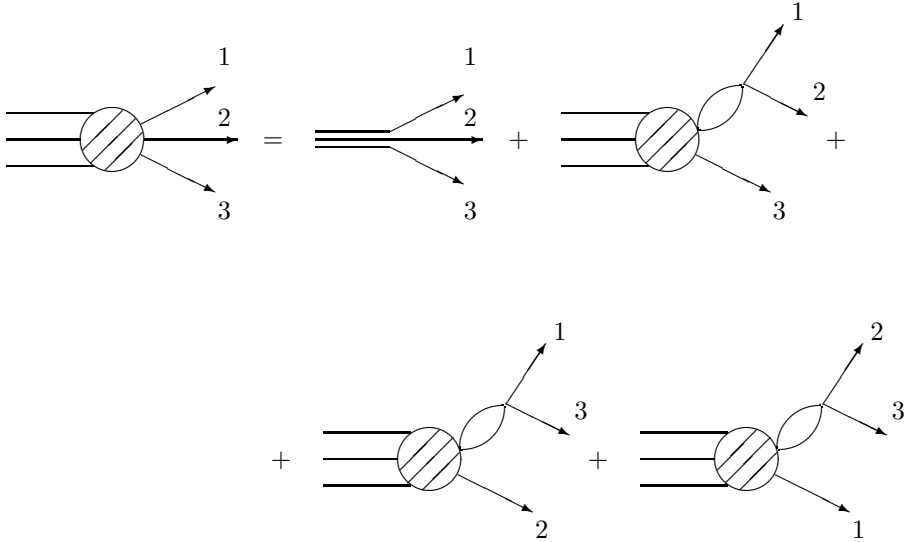
\put(-10,0){{\large Fig.1. The contribution of diagrams at the last pair
of the interacting particles.}}
\end{picture}

\newpage

\vskip60pt
\begin{picture}(600,90)
\put(-10,40){\line(1,0){33}}
\put(-10,50){\line(1,0){28}}
\put(-10,60){\line(1,0){33}}
\put(19,46){\line(1,1){15}}
\put(22,41){\line(1,1){17}}
\put(27.5,38.5){\line(1,1){14}}
\put(41,44){\vector(2,-1){28}}
\put(30,50){\circle{25}}
\put(58,54){\oval(33,33)[tl]}
\put(42,70){\oval(33,33)[br]}
\put(59,71){\vector(2,3){15}}
\put(59,71){\vector(2,-1){24}}
\put(77,95){$1$}
\put(85,65){$2$}
\put(70,20){$3$}
\put(90,47){$=$}
\put(110,52){\line(1,0){28}}
\put(110,50){\line(1,0){28}}
\put(110,48){\line(1,0){28}}
\put(154,51){\oval(33,33)[tl]}
\put(138,67){\oval(33,33)[br]}
\put(155,67){\vector(2,3){15}}
\put(155,67){\vector(2,-1){24}}
\put(139,49){\vector(2,-1){28}}
\put(173,91){$1$}
\put(181,61){$2$}
\put(168,25){$3$}
\put(190,47){$+$}
\put(210,40){\line(1,0){33}}
\put(210,50){\line(1,0){28}}
\put(210,60){\line(1,0){33}}
\put(239,46){\line(1,1){15}}
\put(242,41){\line(1,1){17}}
\put(247.5,38.5){\line(1,1){14}}
\put(261,44){\vector(1,0){43}}
\put(250,50){\circle{25}}
\put(278,54){\oval(33,33)[tl]}
\put(262,70){\oval(33,33)[br]}
\put(279,71){\vector(2,3){15}}
\put(279,71){\vector(1,-1){25}}
\put(297,95){$3$}
\put(300,60){$1$}
\put(290,27){$2$}
\put(305,29){\oval(33,33)[tr]}
\put(321,45){\oval(33,33)[bl]}
\put(323,29){\vector(2,3){15}}
\put(323,29){\vector(2,-1){24}}
\put(341,53){$1$}
\put(333,7){$2$}
\put(360,47){$+$}
\end{picture}

\vskip60pt
\begin{picture}(600,60)
\put(190,47){$+$}
\put(210,40){\line(1,0){33}}
\put(210,50){\line(1,0){28}}
\put(210,60){\line(1,0){33}}
\put(239,46){\line(1,1){15}}
\put(242,41){\line(1,1){17}}
\put(247.5,38.5){\line(1,1){14}}
\put(261,44){\vector(1,0){43}}
\put(250,50){\circle{25}}
\put(278,54){\oval(33,33)[tl]}
\put(262,70){\oval(33,33)[br]}
\put(279,71){\vector(2,3){15}}
\put(279,71){\vector(1,-1){25}}
\put(297,95){$3$}
\put(300,60){$2$}
\put(290,27){$1$}
\put(305,29){\oval(33,33)[tr]}
\put(321,45){\oval(33,33)[bl]}
\put(323,29){\vector(2,3){15}}
\put(323,29){\vector(2,-1){24}}
\put(341,53){$2$}
\put(333,7){$1$}
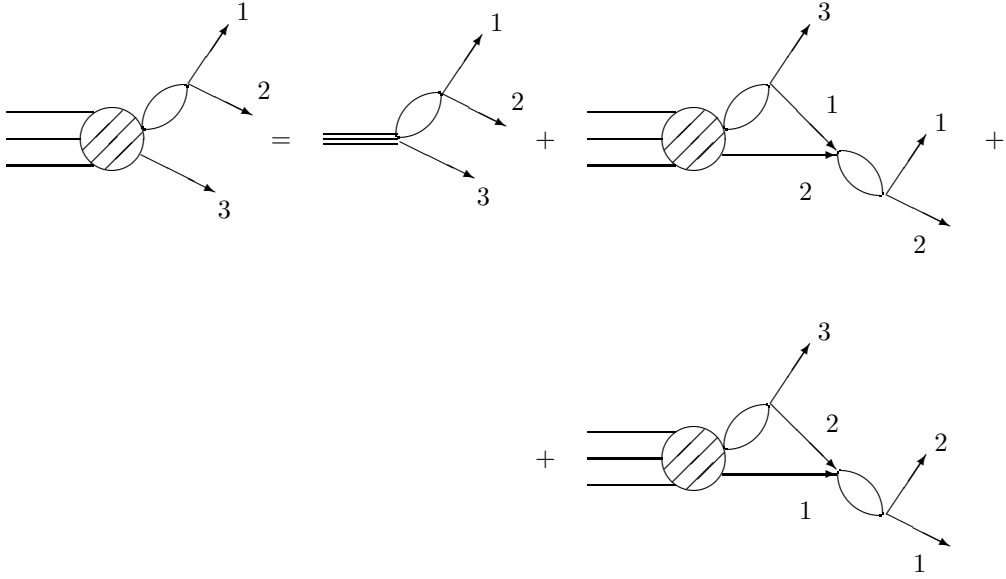
\put(-10,-10){{\large Fig.2. Graphic representation of the equations
for the amplitude $A_1(s,s_{ik})$.}}
\end{picture}

\vskip80pt
\begin{picture}(600,80)
\multiput(70,20)(0,5){20}{\line(0,1){2}}
\put(-10,50){\line(1,0){33}}
\put(-10,60){\line(1,0){28}}
\put(-10,70){\line(1,0){33}}
\put(19,56){\line(1,1){15}}
\put(22,51){\line(1,1){17}}
\put(27.5,48.5){\line(1,1){14}}
\put(41,54){\vector(1,0){43}}
\put(30,60){\circle{25}}
\put(58,64){\oval(33,33)[tl]}
\put(42,80){\oval(33,33)[br]}
\put(59,81){\vector(2,3){15}}
\put(59,81){\vector(1,-1){25}}
\put(77,105){$3$}
\put(75,70){$1, 2$}
\put(60,37){$2, 1$}
\put(90,90){$k_{13}, k_{23}$}
\put(85,39){\oval(33,33)[tr]}
\put(101,55){\oval(33,33)[bl]}
\put(103,39){\vector(2,3){15}}
\put(103,39){\vector(2,-1){24}}
\put(121,40){$k_{12}\frac{A_1^{sym}+3A_0^{asym}}{4}
\left|_{k_{12}}\right.$}
\put(200,70){$+$}
\multiput(305,20)(0,5){20}{\line(0,1){2}}
\put(225,50){\line(1,0){33}}
\put(225,60){\line(1,0){28}}
\put(225,70){\line(1,0){33}}
\put(254,56){\line(1,1){15}}
\put(257,51){\line(1,1){17}}
\put(262.5,48.5){\line(1,1){14}}
\put(276,54){\vector(1,0){43}}
\put(265,60){\circle{25}}
\put(293,64){\oval(33,33)[tl]}
\put(277,80){\oval(33,33)[br]}
\put(294,81){\vector(2,3){15}}
\put(294,81){\vector(1,-1){25}}
\put(312,105){$2$}
\put(310,70){$1, 3$}
\put(295,37){$3, 1$}
\put(325,90){$k_{12}, k_{23}$}
\put(320,39){\oval(33,33)[tr]}
\put(336,55){\oval(33,33)[bl]}
\put(338,39){\vector(2,3){15}}
\put(338,39){\vector(2,-1){24}}
\put(356,40){$k_{13}\frac{A_1^{sym\, s}+3A_0^{asym\, s}}{4}
\left|_{k_{13}}\right.$}
\put(430,70){$+$}
\end{picture}

\vskip60pt
\begin{picture}(600,80)
\multiput(170,20)(0,5){20}{\line(0,1){2}}
\put(65,70){$+$}
\put(90,50){\line(1,0){33}}
\put(90,60){\line(1,0){28}}
\put(90,70){\line(1,0){33}}
\put(119,56){\line(1,1){15}}
\put(122,51){\line(1,1){17}}
\put(127.5,48.5){\line(1,1){14}}
\put(141,54){\vector(1,0){43}}
\put(130,60){\circle{25}}
\put(158,64){\oval(33,33)[tl]}
\put(142,80){\oval(33,33)[br]}
\put(159,81){\vector(2,3){15}}
\put(159,81){\vector(1,-1){25}}
\put(177,105){$1$}
\put(175,70){$2, 3$}
\put(160,37){$3, 2$}
\put(190,90){$k_{12}, k_{13}$}
\put(185,39){\oval(33,33)[tr]}
\put(201,55){\oval(33,33)[bl]}
\put(203,39){\vector(2,3){15}}
\put(203,39){\vector(2,-1){24}}
\put(221,40){$k_{23}A_1^{sym\, s}\left|_{k_{23}}\right.$}
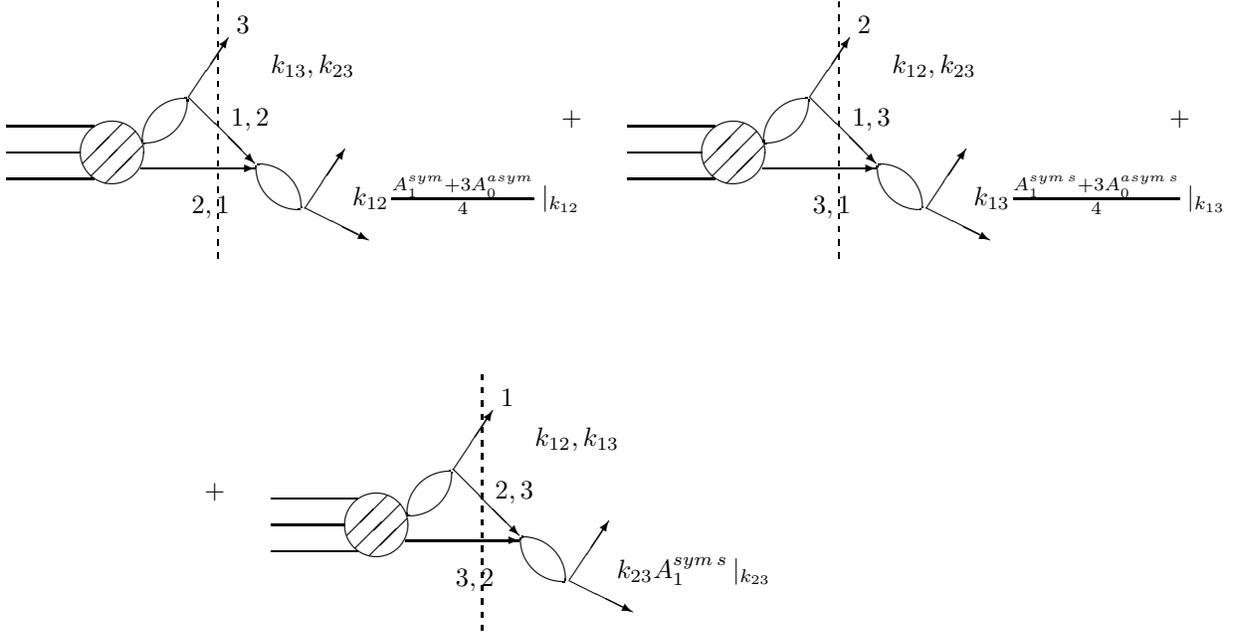
\put(-10,0){{\large Fig.3. The contribution of the diagrams with the
rescattering.}}
\end{picture}

\vskip20pt

We use the diquark projectors. Then we consider the particle $\Sigma$
$\frac{5}{2} ^{+}$ of the $(10,2)$ $(70,2^+)$ multiplet again. This wave
function contains the contribution to $u^2\downarrow u\uparrow s\uparrow$,
which includes three diquarks: $u^2\downarrow u\uparrow$,\,\,
$u^2\downarrow s\uparrow$ \,\, and \,\, $u\uparrow s\uparrow$.
The diquark projectors allow us to obtain the equations (25) -- (27)
(with the definition $\rho_J(s_{ij})\equiv k_{ij}$).

$$k_{12}\left(\frac{A_1^{s0+}+A_1^{p2+}+
2\cdot A_0^{d2+}+2\cdot A_0^{p1-}}{16}
\left(u^2\downarrow u\uparrow s\uparrow
+u\uparrow u^2\downarrow s\uparrow\right)+\right.$$

$$\left. +\frac{A_1^{s0+}+A_1^{p2+}-
2\cdot A_0^{d2+}-2\cdot A_0^{p1-}}{16}\left(
u^2\uparrow u\downarrow s\uparrow+u\downarrow u^2\uparrow s\uparrow\right)
\right.$$

$$\left.
+\frac{A_1^{s0+}+A_1^{p2+}}{8}\left(
u^1\downarrow u^1\uparrow s\uparrow+u^1\uparrow u^1\downarrow s\uparrow
\right)\ \right)\,\, ,\eqno (25)$$

$$k_{13}\left(\frac{A_1^{s0s+}+A_1^{p2s+}+2\cdot A_0^{d2+}+2\cdot A_0^{p1-}
+A_0^{s0s+}+A_0^{p2s+}+2\cdot A_1^{d2+}+2\cdot A_1^{p1-}}{32}
\left(u^2\downarrow u\uparrow s\uparrow
+s\uparrow u\uparrow u^2\downarrow\right)+\right.$$

$$\left. +\frac{A_1^{s0s+}+A_1^{p2s+}-2\cdot A_0^{d2+}-2\cdot A_0^{p1-}
-A_0^{s0s+}-A_0^{p2s+}+2\cdot A_1^{d2+}+2\cdot A_1^{p1-}}{32}
\left(u^2\uparrow u\uparrow s\downarrow
+s\downarrow u\uparrow u^2\uparrow\right)+\right.$$

$$\left. +\frac{A_1^{s0s+}+A_1^{p2s+}+2\cdot A_0^{d2+}+2\cdot A_0^{p1-}
-A_0^{s0s+}-A_0^{p2s+}-2\cdot A_1^{d2+}-2\cdot A_1^{p1-}}{32}
\left(s^2\downarrow u\uparrow u\uparrow
+u\uparrow u\uparrow s^2\downarrow\right)+\right.$$

$$\left. +\frac{A_1^{s0s+}+A_1^{p2s+}-2\cdot A_0^{d2+}-2\cdot A_0^{p1-}
+A_0^{s0s+}+A_0^{p2s+}-2\cdot A_1^{d2+}-2\cdot A_1^{p1-}}{32}
\left(s^2\uparrow u\uparrow u\downarrow
+u\downarrow u\uparrow s^2\uparrow\right)+\right.$$

$$\left. +\frac{A_1^{s0s+}+A_1^{p2s+}+A_0^{s0s+}+A_0^{p2s+}}{16}
\left(u^1\downarrow u\uparrow s^1\uparrow
+s^1\uparrow u\uparrow u^1\downarrow\right)+\right.$$

$$\left. +\frac{A_1^{s0s+}+A_1^{p2s+}-A_0^{s0s+}-A_0^{p2s+}}{16}
\left(u^1\uparrow u\uparrow s^1\downarrow
+s^1\downarrow u\uparrow u^1\uparrow\right)
\right)\,\, ,\eqno (26)$$

$$k_{23}\left(\frac{A_1^{s0s+}+A_1^{p2s+}}{4}
\left(u^2\downarrow u\uparrow s\uparrow
+u^2\downarrow s\uparrow u\uparrow\right)
\right)\,\, .\eqno (27)$$

Then all members of wave function can be considered. After the
groupping of these members we can obtain:

$$u^2\downarrow u\uparrow s\uparrow \left\{
k_{12}\frac{A_1^{s0+}+A_1^{p2+}+3A_0^{d2+}+3A_0^{p1-}}{8}
+k_{13}\frac{A_1^{s0s+}+A_1^{p2s+}+3A_0^{d2s+}+3A_0^{p1s-}}{8}
+k_{23}\, \frac{A_1^{sos+}+A_1^{p2s+}}{2}\right\}\,\, .\eqno (28)$$

The left side of the diagram (Fig.2) corresponds to the
quark interactions. The right side of Fig.2
determines the zero approximation (first diagram) and the subsequent
pair interactions (second diagram). The contribution to
$u^2\downarrow u\uparrow s\uparrow$ is shown in the Fig.3.
If we group the same members, we obtain the system integral equations
for the $\Sigma$ state with the $J^p=\frac{5}{2} ^{+}$ of the $(10,2)$
$(70,2^+)$ multiplet:

$$
\begin{array}{l}
A_1^{s0+}(s,s_{12})=\lambda\, b_{1^{s+}}(s_{12})L_{1^{s+}}(s_{12})+
K_{1^{s+}}(s_{12})\left[\frac{1}{8} A_1^{s0s+}(s,s_{13})
+\frac{1}{8} A_1^{p2s+}(s,s_{13})
+\frac{3}{8} A_0^{d2s+}(s,s_{13})\right.\\
\\
\hskip12.5ex \left.
+\frac{3}{8} A_0^{p1s-}(s,s_{13})
+\frac{1}{8} A_1^{s0s+}(s,s_{23})
+\frac{1}{8} A_1^{p2s+}(s,s_{23})
+\frac{3}{8} A_0^{d2s+}(s,s_{23})
+\frac{3}{8} A_0^{p1s-}(s,s_{23})
\right]\\
\\
A_1^{p2+}(s,s_{12})=\lambda\, b_{3^{d+}}(s_{12})L_{3^{d+}}(s_{12})+
K_{3^{d+}}(s_{12})\left[\frac{1}{8} A_1^{s0s+}(s,s_{13})
+\frac{1}{8} A_1^{p2s+}(s,s_{13})
+\frac{3}{8} A_0^{d2s+}(s,s_{13})\right.\\
\\
\hskip12.5ex \left.
+\frac{3}{8} A_0^{p1s+}(s,s_{13})
+\frac{1}{8} A_1^{s0s+}(s,s_{23})
+\frac{1}{8} A_1^{p2s+}(s,s_{23})
+\frac{3}{8} A_0^{d2s+}(s,s_{23})
+\frac{3}{8} A_0^{p1s-}(s,s_{23})
\right]\\
\\
A_1^{s0s+}(s,s_{12})=
\lambda\, b_{1^{s+}_s}(s_{12})L_{1^{s+}_s}(s_{12})+
K_{1^{s+}_s}(s_{12})\left[\frac{1}{4} A_1^{s0+}(s,s_{13})
+\frac{1}{4} A_1^{p2+}(s,s_{13})
-\frac{1}{8} A_1^{s0s+}(s,s_{13})\right.\\
\\
\hskip13ex \left.
-\frac{1}{8} A_1^{p2s+}(s,s_{13})
+\frac{3}{8} A_0^{d2s+}(s,s_{13})
+\frac{3}{8} A_0^{p1s-}(s,s_{13})
+\frac{1}{4} A_1^{s0+}(s,s_{23})
+\frac{1}{4} A_1^{p2+}(s,s_{23})\right.\\
\\
\hskip13ex \left.
-\frac{1}{8} A_1^{s0s+}(s,s_{23})
-\frac{1}{8} A_1^{p2s+}(s,s_{23})
+\frac{3}{8} A_0^{d2s+}(s,s_{23})
+\frac{3}{8} A_0^{p1s-}(s,s_{23})
\right]\\
\end{array}$$

$$\begin{array}{l}
A_1^{p2s+}(s,s_{12})=
\lambda\, b_{3^{d+}_s}(s_{12})L_{3^{d+}_s}(s_{12})+
K_{3^{d+}_s}(s_{12})\left[\frac{1}{4} A_1^{s0+}(s,s_{13})
+\frac{1}{4} A_1^{p2+}(s,s_{13})
-\frac{1}{8} A_1^{s0s+}(s,s_{13})\right.\\
\\
\hskip13ex \left.
-\frac{1}{8} A_1^{p2s+}(s,s_{13})
+\frac{3}{8} A_0^{d2s+}(s,s_{13})
+\frac{3}{8} A_0^{p1s-}(s,s_{13})
+\frac{1}{4} A_1^{s0+}(s,s_{23})
+\frac{1}{4} A_1^{p2+}(s,s_{23})\right.\\
\\
\hskip13ex \left.
-\frac{1}{8} A_1^{s0s+}(s,s_{23})
-\frac{1}{8} A_1^{p2s+}(s,s_{23})
+\frac{3}{8} A_0^{d2s+}(s,s_{23})
+\frac{3}{8} A_0^{p1s-}(s,s_{23})
\right]\\
\\
A_0^{d2s+}(s,s_{12})=
\lambda\, b_{2^{d+}_s}(s_{12})L_{2^{d+}_s}(s_{12})+
K_{2^{d+}_s}(s_{12})\left[\frac{1}{4} A_1^{s0+}(s,s_{13})
+\frac{1}{4} A_1^{p2+}(s,s_{13})
+\frac{1}{8} A_1^{s0s+}(s,s_{13})\right.\\
\\
\hskip13ex \left.
+\frac{1}{8} A_1^{p2s+}(s,s_{13})
+\frac{1}{8} A_0^{d2s+}(s,s_{13})
+\frac{1}{8} A_0^{p1s-}(s,s_{13})
+\frac{1}{4} A_1^{s0+}(s,s_{23})
+\frac{1}{4} A_1^{p2+}(s,s_{23})\right.\\
\\
\hskip13ex \left.
+\frac{1}{8} A_1^{s0s+}(s,s_{23})
+\frac{1}{8} A_1^{p2s+}(s,s_{23})
+\frac{1}{8} A_0^{d2s+}(s,s_{23})
+\frac{1}{8} A_0^{p1s-}(s,s_{23})
\right]\\
\\
A_0^{p1s-}(s,s_{12})=
\lambda\, b_{1^{p-}_s}(s_{12})L_{1^{p-}_s}(s_{12})+
K_{1^{p-}_s}(s_{12})\left[\frac{1}{4} A_1^{s0+}(s,s_{13})
+\frac{1}{4} A_1^{p2+}(s,s_{13})
+\frac{1}{8} A_1^{s0s+}(s,s_{13})\right.\\
\\
\hskip13ex \left.
+\frac{1}{8} A_1^{p2s+}(s,s_{13})
+\frac{1}{8} A_0^{d2s+}(s,s_{13})
+\frac{1}{8} A_0^{p1s-}(s,s_{13})
+\frac{1}{4} A_1^{s0+}(s,s_{23})
+\frac{1}{4} A_1^{p2+}(s,s_{23})\right.\\
\\
\hskip13ex \left.
+\frac{1}{8} A_1^{s0s+}(s,s_{23})
+\frac{1}{8} A_1^{p2s+}(s,s_{23})
+\frac{1}{8} A_0^{d2s+}(s,s_{23})
+\frac{1}{8} A_0^{p1s-}(s,s_{23})
\right] .\\
\end{array}
\eqno (29)$$

Here the function $L_J(s_{ik})$ has the form

$$L_J(s_{ik})=\frac{G_J(s_{ik})}{1-B_J(s_{ik})}.\eqno (30)$$

The integral operator $K_J (s_{ik})$ is:

$$K_J (s_{ik})=L_J(s_{ik})\, \int\limits_{(m_i+m_k)^2}^{\infty}\hskip2mm
\frac{ds'_{ik}}{\pi}\frac{\rho_J(s'_{ik})G_J(s'_{ik})}
{s'_{ik}-s_{ik}}\, \int\limits_{-1}^{1}\frac{dz}{2}\, ,\eqno (31)$$

$$b_J(s_{ik})=\int\limits_{(m_i+m_k)^2}^{\Lambda_J(i,k)}
\frac{ds'_{ik}}{\pi}\frac{\rho_J(s'_{ik})G^2_J(s'_{ik})}
{s'_{ik}-s_{ik}}\, .\eqno (32)$$

The function $b_J(s_{ik})$ is the truncate function of Chew-Mandelstam.
$z$ is the cosine of the angle between the relative momentum of particles
$i$ and $k$ in the intermediate state and the momentum of
particle $j$ in the final state, taken in the c.m. of the particles
$i$ and $k$. $\lambda$ is the current constant. By analogy with the
$\Sigma$ $\frac{5}{2} ^{+}$ $(10,2)$ $(70,2^+)$ state we obtain
the rescattering amplitudes of the three various quarks for all $(70,2^+)$
and $(70,0^+)$ states which satisfy the system of integral equations
(Appendix I).

\vskip2.0ex
{\large \bf 4. The reduced equations for the $(70,0^+)$
and $(70,2^+)$ multiplets.}
\vskip2.0ex

Let us extract two-particle singularities in $A_J(s,s_{ik})$:

$$A_J(s,s_{ik})=\frac{\alpha_J(s,s_{ik})b_J(s_{ik})G_J(s_{ik})}
{1-B_J(s_{ik})},\eqno (33)$$

\noindent
$\alpha_J(s,s_{ik})$ is the reduced amplitude. Accordingly to this,
all integral equations can be rewritten using the reduced
amplitudes. For instance, one considers the first equation of system
for the $\Sigma$ $J^p=\frac{5}{2}^+$ of the $(10,2)$ $(70,2^+)$ multiplet:

$$\begin{array}{l}
\alpha_1^{s0+} (s,s_{12})=\lambda+\frac{1}{b_{1^{s+}}(s_{12})}
\, \int\limits_{(m_1+m_2)^2}^{\Lambda_{1^+}(1,2)}\hskip2mm
\frac{ds'_{12}}{\pi}\,\frac{\rho_{1^{s+}}(s'_{12})G_{1^{s+}}(s'_{12})}
{s'_{12}-s_{12}}\\
\\
\hskip13ex
\times \int\limits_{-1}^{1}\frac{dz}{2}\,
\left(
\frac{G_{1_s^{s+}}(s'_{13})b_{1_s^{s+}}(s'_{13})}{1-B_{1_s^{s+}}(s'_{13})}
\,\frac{1}{4}\,\alpha_1^{s0s+}(s,s'_{13})
+\frac{G_{3_s^{d+}}(s'_{13})b_{3_s^{d+}}(s'_{13})}{1-B_{3_s^{d+}}(s'_{13})}
\,\frac{1}{4}\,\alpha_1^{p2s+}(s,s'_{13})\right.\\
\\
\hskip13ex \left.
+\frac{G_{2_s^{d+}}(s'_{13})b_{2_s^{d+}}(s'_{13})}{1-B_{2_s^{d+}}(s'_{13})}
\,\frac{3}{4}\,\alpha_0^{d2s+}(s,s'_{13})
+\frac{G_{1_s^{p-}}(s'_{13})b_{1_s^{p-}}(s'_{13})}{1-B_{1_s^{p-}}(s'_{13})}
\,\frac{3}{4}\,\alpha_0^{p1s-}(s,s'_{13})
\right).\end{array}\eqno (34)$$

The connection between $s'_{12}$ and $s'_{13}$ is:

$$s'_{13}=m_1^2+m_3^2-\frac{\left(s'_{12}+m_3^2-s\right)
\left(s'_{12}+m_1^2-m_2^2\right)}{2s'_{12}}\pm$$

$$\pm\frac{z}{2s'_{12}}\times\sqrt{\left(s'_{12}-(m_1+m_2)^2\right)
\left(s'_{12}-(m_1-m_2)^2\right)}\times$$

$$\times\sqrt{\left(s'_{12}-(\sqrt{s}+m_3)^2\right)
\left(s'_{12}-(\sqrt{s}-m_3)^2\right)}\, .\eqno (35)$$

The formula for $s'_{23}$ is similar to (35) with $z$ replaced by $-z$.
Thus $A_1^{s0s+}(s,s'_{13})+A_1^{s0s+}(s,s'_{23})$ must be replaced by
$2A_1^{s0s+}(s,s'_{13})$. $\Lambda_J(i,k)$ is the cutoff at the large
value of $s_{ik}$, which determines the contribution from small distances.

The construction of the approximate solution of system of equations is
based on the extraction of the leading singularities which are close to the
region $s_{ik}=(m_i+m_k)^2$ [29]. Amplitudes with different number of
rescattering have the following structure of singularities. The main
singularities in $s_{ik}$ are from pair rescattering of the particles
$i$ and $k$. First of all there are threshold square root singularities.
Pole singularities are also possible which correspond to the bound
states. The diagrams in Fig.2 apart from two-particle singularities
have their own specific triangle singularities. Such classification
allows us to search the approximate solution of by taking into
account some definite number of leading singularities and neglecting all
the weaker ones.

We consider the approximation, which corresponds to the single interaction
of all three particles (two-particle and triangle singularities) and
neglecting all the weaker ones.

The functions $\alpha_J(s,s_{ik})$ are the smooth functions of $s_{ik}$
as compared with the singular part of the amplitude, hence it can be
expanded in a series in the singulary point and only the first term of
this series should be employed further. As $s_0$ it is convenient to
take the middle point of physical region of Dalitz-plot in which $z=0$.
In this case we get from (35)
$s_{ik}=s_0=\frac{s+m_1^2+m_2^2+m_3^2}{m_{12}^2+m_{13}^2+m_{23}^2}$,
where $m_{ik}=\frac{m_i+m_k}{2}$. We define the functions
$\alpha_J(s,s_{ik})$ and $b_J(s_{ik})$ at the point $s_0$. Such a choice
of point $s_0$ allows us to replace integral equations (29) by the
algebraic equations for the state $\Sigma$ with $J^p=\frac{5}{2}^+$
of $(10,2)$ $(70,2^+)$:

$$\left\{\begin{array}{l}
\alpha_1^{s0+}=\lambda+
\frac{1}{4}\,\,\alpha_1^{s0s+}\,\, M_{1^{s+} 1^{s+}_s}
+\frac{1}{4}\,\,\alpha_1^{p2s+}\,\, M_{1^{s+} 3^{d+}_s}
+\frac{3}{4}\,\,\alpha_0^{d2s+}\,\, M_{1^{s+} 2^{d+}_s}
+\frac{3}{4}\,\,\alpha_0^{p1s-}\,\, M_{1^{s+} 1^{p-}_s}
\hskip4.0ex 1^{s+}\\
\\
\alpha_1^{p2+}=\lambda+
\frac{1}{4}\,\,\alpha_1^{s0s+}\,\, M_{3^{d+} 1^{s+}_s}
+\frac{1}{4}\,\,\alpha_1^{p2s+}\,\, M_{3^{d+} 3^{d+}_s}
+\frac{3}{4}\,\,\alpha_0^{d2s+}\,\, M_{3^{d+} 2^{d+}_s}
+\frac{3}{4}\,\,\alpha_0^{p1s-}\,\, M_{3^{d+} 1^{p-}_s}
\hskip4.0ex 3^{d+}\\
\\
\alpha_1^{s0s+}=\lambda+
\frac{1}{2}\,\,\alpha_1^{s0+}\,\, M_{1^{s+}_s 1^{s+}}
+\frac{1}{2}\,\,\alpha_1^{p2+}\,\, M_{1^{s+}_s 3^{d+}}
-\frac{1}{4}\,\,\alpha_1^{s0s+}\,\, M_{1^{s+}_s 1^{s+}_s}
-\frac{1}{4}\,\,\alpha_1^{p2s+}\,\,M_{1^{s+}_s 3^{d+}_s}
\hskip5.5ex 1^{s+}_s\\
\\
\hskip6ex
+\frac{3}{4}\,\,\alpha_0^{d2s+}\,\,M_{1^{s+}_s 2^{d+}_s}
+\frac{3}{4}\,\,\alpha_0^{p1s-}\,\,M_{1^{s+}_s 1^{p-}_s}\\
\\
\alpha_1^{p2s+}=\lambda+
\frac{1}{2}\,\,\alpha_1^{s0+}\,\, M_{3^{d+}_s 1^{s+}}
+\frac{1}{2}\,\,\alpha_1^{p2+}\,\, M_{3^{d+}_s 3^{d+}}
-\frac{1}{4}\,\,\alpha_1^{s0s+}\,\, M_{3^{d+}_s 1^{s+}_s}
-\frac{1}{4}\,\,\alpha_1^{p2s+}\,\,M_{3^{d+}_s 3^{d+}_s}
\hskip5.5ex 3^{d+}_s\\
\\
\hskip6ex
+\frac{3}{4}\,\,\alpha_0^{d2s+}\,\,M_{3^{d+}_s 2^{d+}_s}
+\frac{3}{4}\,\,\alpha_0^{p1s-}\,\,M_{3^{d+}_s 1^{p-}_s}\\
\\
\alpha_0^{d2s+}=\lambda+
\frac{1}{2}\,\,\alpha_1^{s0+}\,\, M_{2^{d+}_s 1^{s+}}
+\frac{1}{2}\,\,\alpha_1^{p2+}\,\, M_{2^{d+}_s 3^{d+}}
+\frac{1}{4}\,\,\alpha_1^{s0s+}\,\, M_{2^{d+}_s 1^{s+}_s}
+\frac{1}{4}\,\,\alpha_1^{p2s+}\,\,M_{2^{d+}_s 3^{d+}_s}
\hskip5.5ex 2^{d+}_s\\
\\
\hskip6ex
+\frac{1}{4}\,\,\alpha_0^{d2s+}\,\,M_{2^{d+}_s 2^{d+}_s}
+\frac{1}{4}\,\,\alpha_0^{p1s-}\,\,M_{2^{d+}_s 1^{p-}_s}\\
\\
\alpha_0^{p1s-}=\lambda+
\frac{1}{2}\,\,\alpha_1^{s0+}\,\, M_{1^{p-}_s 1^{s+}}
+\frac{1}{2}\,\,\alpha_1^{p2+}\,\, M_{1^{p-}_s 3^{d+}}
+\frac{1}{4}\,\,\alpha_1^{s0s+}\,\, M_{1^{p-}_s 1^{s+}_s}
+\frac{1}{4}\,\,\alpha_1^{p2s+}\,\,M_{1^{p-}_s 3^{d+}_s}
\hskip5.0ex 1^{p-}_s\\
\\
\hskip6ex
+\frac{1}{4}\,\,\alpha_0^{d2s+}\,\,M_{1^{p-}_s 2^{d+}_s}
+\frac{1}{4}\,\,\alpha_0^{p1s-}\,\,M_{1^{p-}_s 1^{p-}_s}\,\, ,\\
\end{array}\right. \eqno (36)$$

Here the reduced amplitudes for the diquarks $1^{s+}$, $3^{d+}$,
$1^{s+}_s$, $3^{d+}_s$, $2^{d+}_s$, $1^{p-}_s$ are given.

We used the following form:

$$M_{X^{ip}_m Y^{jq}_n}\equiv
M_{X^{ip}_m Y^{jq}_n}(s,s_0)=I_{X^{ip}_m Y^{jq}_n}(s,s_0)
\,\,\frac{b_{Y^{jq}_n}(s_0)}{b_{X^{ip}_m}(s_0)}\,\, ,\eqno (37)$$

\noindent
here $X^{ip}_m$ corresponds to the diquark with total moment $X$
($X=0,1,2,3$); $i=s,p,d$ for the $s$-, $p$-, $d$-wave consequently;
$p=+,-$ is the $p$-parity of diquark; $m=s$ for the strange diquark and
this index is absent in other case.

The reduced amplitude $\alpha_s^{clmp}\equiv \alpha_s^{clmp} (s,s_0)$
for the $p=+,-$ of parity of diquark; $c=s$ if the diquark is
determined as $1s1s$, $c=p$ if we consider $1s1p$ or $1p1p$ states,
$c=d$ if we have $1s1d$; $s=1,0$ corresponds to the diquark spin
($\uparrow \uparrow$, $\uparrow\downarrow$), $l=2,1,0,-1$ are the values
of projection orbital moment at definite axies, $m=s$ for the strange
diquark.

The function $I_{J_1 J_2}(s,s_0)$ takes into account the singularity which
corresponds to the simultaneous vanishing of all propagators in the
triangle diagrams.

$$I_{J_1 J_2}(s,s_0)=\int\limits_{(m_i+m_k)^2}^{\Lambda_{J_1}}\hskip2mm
\frac{ds'_{ik}}{\pi}\frac{\rho_{J_1}(s'_{ik})G^2_{J_1}(s'_{ik})}
{s'_{ik}-s_{ik}}\, \int\limits_{-1}^{1}\frac{dz}{2}\,
\frac{1}{1-B_{J_2}(s_{ij})}\, .\eqno (38)$$

The $G_J(s_{ik})$ functions have the smooth dependence from energy
$s_{ik}$ [27] therefore we suggest them as constants.
The parameters of model: $g_J$ vertex constants, $\lambda_J$ cutoff
parameters are chosen dimensionless.

$$g_J=\frac{m_i+m_k}{2\pi}G_J , \,\,\, \lambda_J=\frac{4\Lambda_J}
{(m_i+m_k)^2} .\eqno (39)$$

Here $m_i$ and $m_k$ are quark masses in the intermediate state of the quark
loop. Dimensionless parameters $g_J$ and $\lambda_J$ are supposed to be
the constants independent of the quark interaction type. We calculate
the system of equations and can determine the mass values of the
$\Sigma$ $J^p=\frac{5}{2}^+$ $(10,2)$ $(70,2^+)$. We calculate a pole
in $s$ which corresponds to the bound state of the three quarks.

By analogy with the $\Sigma$-hyperon we obtain the system of equations for
the reduced amplitudes for all particles $(70,0^+)$ and $(70,2^+)$
multiplets (Appendix II).

The solutions of the system of equations are considered as:

$$\alpha_J=\frac{F_J(s,\lambda_J)}{D(s)}\, , \eqno(40)$$

\noindent
where the zeros of the $D(s)$ determinates define of masses of bound
states of baryons. $F_J(s,\lambda_J)$ are the functions of $s$ and
$\lambda_J$. The functions $F_J(s,\lambda_J)$ determine the contributions
of subamplitudes to the excited baryon amplitude.

\vskip30pt

\noindent
{\large Table I.}

\noindent
{The $\Delta$-isobar masses of multiplet $(70,2^+)$.}

\vskip1.5ex

\noindent
\begin{tabular}{|c|c|c|c|}
\hline
Multiplet & Baryon & Mass ($GeV$) & Mass ($GeV$) (exp.) \\
\hline
$\frac{5}{2}^+$ $(10,2)$ & $F_{35}$ & $2.000$ & $2.000$\\
\hline
$\frac{3}{2}^+$ $(10,2)$ & $P_{33}$ & $2.088$ & $-$\\
\hline
\end{tabular}

\vskip1.5ex

\noindent
{The parameters of model (Tables I-XII): gluon coupling constants
$g^+_s =g^-_p =0.739$, $g^+_d =0.550$,\\ cutoff energy
parameters $\lambda=10.0$, $\lambda_{ss}=8.9$.}

\vskip30pt

\noindent
{\large Table II.}

\noindent
{The nucleon masses of multiplet $(70,2^+)$.}

\vskip1.5ex

\noindent
\begin{tabular}{|c|c|c|c|}
\hline
Multiplet & Baryon & Mass ($GeV$) & Mass ($GeV$) (exp.) \\
\hline
$\frac{5}{2}^+$ $(8,2)$ & $F_{15}$ & $2.000$ & $2.000$\\
\hline
$\frac{3}{2}^+$ $(8,2)$ & $P_{13}$ & $2.045$ & $-$\\
\hline
$\frac{7}{2}^+$ $(8,4)$ & $F_{17}$ & $2.074$ & $1.990$\\
\hline
$\frac{5}{2}^+$ $(8,4)$ & $F_{15}$ & $2.009$ & $-$\\
\hline
$\frac{3}{2}^+$ $(8,4)$ & $P_{13}$ & $1.971$ & $1.900$\\
\hline
$\frac{1}{2}^+$ $(8,4)$ & $P_{11}$ & $1.682$ & $-$\\
\hline
\end{tabular}

\vskip30pt

\noindent
{\large Table III.}

\noindent
{The $\Sigma$-hyperon masses of multiplet $(70,2^+)$.}

\vskip1.5ex

\noindent
\begin{tabular}{|c|c|c|c|}
\hline
Multiplet & Baryon & Mass ($GeV$) & Mass ($GeV$) (exp.) \\
\hline
$\frac{5}{2}^+$ $(10,2)$ & $F_{35}$ & $2.049$ & $-$\\
\hline
$\frac{3}{2}^+$ $(10,2)$ & $P_{33}$ & $2.179$ & $-$\\
\hline
$\frac{5}{2}^+$ $(8,2)$ & $F_{15}$ & $2.057$ & $-$\\
\hline
$\frac{3}{2}^+$ $(8,2)$ & $P_{13}$ & $2.118$ & $2.080$\\
\hline
$\frac{7}{2}^+$ $(8,4)$ & $F_{17}$ & $2.159$ & $2.030$\\
\hline
$\frac{5}{2}^+$ $(8,4)$ & $F_{15}$ & $2.070$ & $2.070$\\
\hline
$\frac{3}{2}^+$ $(8,4)$ & $P_{13}$ & $2.033$ & $-$\\
\hline
$\frac{1}{2}^+$ $(8,4)$ & $P_{11}$ & $1.660$ & $1.660$\\
\hline
\end{tabular}

\vskip30pt

\noindent
{\large Table IV.}

\noindent
{The $\Xi$-hyperon masses of multiplet $(70,2^+)$.}

\vskip1.5ex

\noindent
\begin{tabular}{|c|c|c|c|}
\hline
Multiplet & Baryon & Mass ($GeV$) & Mass ($GeV$) (exp.) \\
\hline
$\frac{5}{2}^+$ $(10,2)$ & $F_{35}$ & $2.135$ & $-$\\
\hline
$\frac{3}{2}^+$ $(10,2)$ & $P_{33}$ & $2.280$ & $-$\\
\hline
$\frac{5}{2}^+$ $(8,2)$ & $F_{15}$ & $2.143$ & $-$\\
\hline
$\frac{3}{2}^+$ $(8,2)$ & $P_{13}$ & $2.212$ & $-$\\
\hline
$\frac{7}{2}^+$ $(8,4)$ & $F_{17}$ & $2.258$ & $-$\\
\hline
$\frac{5}{2}^+$ $(8,4)$ & $F_{15}$ & $2.154$ & $-$\\
\hline
$\frac{3}{2}^+$ $(8,4)$ & $P_{13}$ & $2.109$ & $-$\\
\hline
$\frac{1}{2}^+$ $(8,4)$ & $P_{11}$ & $1.659$ & $-$\\
\hline
\end{tabular}

\vskip30pt

\noindent
{\large Table V.}

\noindent
{The $\Lambda$-hyperon masses of multiplet $(70,2^+)$.}

\vskip1.5ex

\noindent
\begin{tabular}{|c|c|c|c|}
\hline
Multiplet & Baryon & Mass ($GeV$) & Mass ($GeV$) (exp.) \\
\hline
$\frac{5}{2}^+$ $(8,2)$ & $F_{15}$ & $2.123$ & $2.110$\\
\hline
$\frac{3}{2}^+$ $(8,2)$ & $P_{13}$ & $2.061$ & $-$\\
\hline
$\frac{7}{2}^+$ $(8,4)$ & $F_{17}$ & $2.158$ & $-$\\
\hline
$\frac{5}{2}^+$ $(8,4)$ & $F_{15}$ & $2.073$ & $-$\\
\hline
$\frac{3}{2}^+$ $(8,4)$ & $P_{13}$ & $2.022$ & $-$\\
\hline
$\frac{1}{2}^+$ $(8,4)$ & $P_{11}$ & $1.649$ & $-$\\
\hline
$\frac{5}{2}^+$ $(1,2)$ & $F_{05}$ & $2.074$ & $-$\\
\hline
$\frac{3}{2}^+$ $(1,2)$ & $P_{03}$ & $2.056$ & $-$\\
\hline
\end{tabular}

\vskip30pt

\noindent
{\large Table VI.}

\noindent
{The $\Omega$-hyperon masses of multiplet $(70,2^+)$.}

\vskip1.5ex

\noindent
\begin{tabular}{|c|c|c|c|}
\hline
Multiplet & Baryon & Mass ($GeV$) & Mass ($GeV$) (exp.) \\
\hline
$\frac{5}{2}^+$ $(10,2)$ & $F_{35}$ & $2.250$ & $-$\\
\hline
$\frac{3}{2}^+$ $(10,2)$ & $P_{33}$ & $2.406$ & $-$\\
\hline
\end{tabular}

\newpage

\noindent
{\large Table VII.}

\noindent
{The $\Delta$-isobar masses of multiplet $(70,0^+)$.}

\vskip1.5ex

\noindent
\begin{tabular}{|c|c|c|c|}
\hline
Multiplet & Baryon & Mass ($GeV$) & Mass ($GeV$) (exp.) \\
\hline
$\frac{1}{2}^+$ $(10,2)$ & $P_{31}$ & $1.750$ & $1.750$\\
\hline
\end{tabular}

\vskip30pt

\noindent
{\large Table VIII.}

\noindent
{The nucleon masses of multiplet $(70,0^+)$.}

\vskip1.5ex

\noindent
\begin{tabular}{|c|c|c|c|}
\hline
Multiplet & Baryon & Mass ($GeV$) & Mass ($GeV$) (exp.) \\
\hline
$\frac{1}{2}^+$ $(8,2)$ & $P_{11}$ & $1.710$ & $1.710$\\
\hline
$\frac{3}{2}^+$ $(8,4)$ & $P_{13}$ & $1.791$ & $-$\\
\hline
\end{tabular}

\vskip30pt

\noindent
{\large Table IX.}

\noindent
{The $\Sigma$-hyperon masses of multiplet $(70,0^+)$.}

\vskip1.5ex

\noindent
\begin{tabular}{|c|c|c|c|}
\hline
Multiplet & Baryon & Mass ($GeV$) & Mass ($GeV$) (exp.) \\
\hline
$\frac{1}{2}^+$ $(10,2)$ & $P_{31}$ & $1.740$ & $-$\\
\hline
$\frac{1}{2}^+$ $(8,2)$ & $P_{11}$ & $1.677$ & $1.770$\\
\hline
$\frac{3}{2}^+$ $(8,4)$ & $P_{13}$ & $1.799$ & $-$\\
\hline
\end{tabular}

\vskip30pt

\noindent
{\large Table X.}

\noindent
{The $\Xi$-hyperon masses of multiplet $(70,0^+)$.}

\vskip1.5ex

\noindent
\begin{tabular}{|c|c|c|c|}
\hline
Multiplet & Baryon & Mass ($GeV$) & Mass ($GeV$) (exp.) \\
\hline
$\frac{1}{2}^+$ $(10,2)$ & $P_{31}$ & $1.774$ & $-$\\
\hline
$\frac{1}{2}^+$ $(8,2)$ & $P_{11}$ & $1.701$ & $-$\\
\hline
$\frac{3}{2}^+$ $(8,4)$ & $P_{13}$ & $1.840$ & $-$\\
\hline
\end{tabular}

\vskip30pt

\noindent
{\large Table XI.}

\noindent
{The $\Lambda$-hyperon masses of multiplet $(70,0^+)$.}

\vskip1.5ex

\noindent
\begin{tabular}{|c|c|c|c|}
\hline
Multiplet & Baryon & Mass ($GeV$) & Mass ($GeV$) (exp.) \\
\hline
$\frac{1}{2}^+$ $(8,2)$ & $P_{11}$ & $1.721$ & $1.810$\\
\hline
$\frac{3}{2}^+$ $(8,4)$ & $P_{13}$ & $1.800$ & $1.890$\\
\hline
$\frac{1}{2}^+$ $(1,2)$ & $P_{01}$ & $1.627$ & $1.600$\\
\hline
\end{tabular}

\vskip30pt

\noindent
{\large Table XII.}

\noindent
{The $\Omega$-hyperon masses of multiplet $(70,0^+)$.}

\vskip1.5ex

\noindent
\begin{tabular}{|c|c|c|c|}
\hline
Multiplet & Baryon & Mass ($GeV$) & Mass ($GeV$) (exp.) \\
\hline
$\frac{1}{2}^+$ $(10,2)$ & $P_{31}$ & $1.865$ & $-$\\
\hline
\end{tabular}

\newpage

\noindent
{\large Table XIII.}

\noindent
{Coefficient of Ghew-Mandelstam function for the
different diquarks.}

\vskip1.5ex

\noindent
\begin{tabular}{|c|c|c|c|}
\hline
 &$\alpha_J$&$\beta_J$&$\delta_J$\\
\hline
 & & & \\
$3^+$&$\frac{5}{14}$&
$\frac{2}{14}-\frac{5}{14}\frac{(m_i-m_k)^2}{(m_i+m_k)^2}$
&$-\frac{2}{14}(m_i-m_k)^2$\\
 & & & \\
$2^+$&$\frac{1}{2}$&$-\frac{1}{2}\frac{(m_i-m_k)^2}{(m_i+m_k)^2}$
&$0$\\
 & & & \\
$1^{d+}$&$\frac{1}{7}$&$
\frac{3}{14}\left(1+\frac{8m_i m_k}{3(m_i+m_k)^2}\right)$
&$-\frac{5}{14}(m_i-m_k)^2$\\
 & & & \\
$1^{s+}$&$\frac{1}{3}$&$\frac{4m_i m_k}{3(m_i+m_k)^2}-\frac{1}{6}$
&$-\frac{1}{6}(m_i-m_k)^2$\\
 & & & \\
$0^+$&$\frac{1}{2}$&$-\frac{1}{2}\frac{(m_i-m_k)^2}{(m_i+m_k)^2}$&0\\
 & & & \\
$0^-$&$0$&$\frac{1}{2}$&$-\frac{1}{2}(m_i-m_k)^2$\\
 & & & \\
$1^-$&$\frac{1}{2}$&$-\frac{1}{2}\frac{(m_i-m_k)^2}{(m_i+m_k)^2}$&0\\
 & & & \\
$2^-$&$\frac{3}{10}$&$\frac{1}{5}
\left(1-\frac{3}{2}\frac{(m_i-m_k)^2}{(m_i+m_k)^2}\right)$
&$-\frac{1}{5}(m_i-m_k)^2$\\
 & & & \\
\hline
\end{tabular}

\vskip2.0ex
{\large \bf 5. Calculation results.}
\vskip2.0ex

The quark masses ($m_u=m_d=m$ and $m_s$) are not fixed. In order to fix
$m$ and $m_s$, in any way we assume $m=\frac{1}{3}m_{\Delta}(1.232)$ and
$m=\frac{1}{3}m_{\Omega}(1.672)$ i.e. the quark masses are $m=0.410\, GeV$
and $m_s=0.557\, GeV$.

The $S$-wave baryon mass spectra are obtained in good agreement with the
experimental data. When we research the excited states the
confinement potential can not be neglected. In our case the confinement
potential is imitated by the simple increasing of constituent quark
masses [30]. The shift of quark mass (parameter $\Delta=340\, MeV$)
effectively takes into account the changing of the confinement potential.
We have shown that inclusion of only gluon exchange does not lead
to the appearance of bound states corresponding to the excited baryons
in the $(70,0^+)$ and $(70,2^+)$ multiplets. The mass shift $\Delta$
allows to obtain the mass spectra of these states.
The similar result for the $P$-wave baryons was obtained [24].

In the case considered the same parameters $\Delta$ for the
$u, d, s$ quarks are chosen. Then the quark masses $m_u=m_d=0.750\, GeV$
and $m_s=0.897\, GeV$ are given.

In our model the four parameters are used: gluon coupling constants
$g_s^+\equiv g_p^-$ for the $s$- and $p$-wave diquarks, $g_d^+$ for
$d$-wave diquarks, cutoff parameters $\lambda$, $\lambda _{ss}$ for the
nonstrange and strange diquarks. Parameter of $\lambda _s$ was calculated
using $\lambda$ and $\lambda _{ss}$ parameters.

The parameters $g_s^+\equiv g_p^-=0.739$, $g_d+=0.550$, $\lambda=10.0$,
$\lambda_{ss}=8.9$ have been determined by the baryon masses:
$M_{\Delta \frac{1}{2}^+ (10,2)}=1.750\, GeV$,\,
$M_{N \frac{1}{2}^+ (8,2)}=1.710\, GeV$,\,
$M_{\Delta \frac{5}{2}^+ (10,2)}=2.000\, GeV$, and
$M_{\Sigma \frac{5}{2}^+ (8,4)}=2.070\, GeV$.

In the tables I-XII we represent the masses of the nonstrange and strange
resonances belonging to the $(70,0^+)$ and $(70,2^+)$ multiplets obtained
using the fit of the experimental values [25].

The $(70,0^+)$ and $(70,2^+)$ multiplets include $414$ particles, only $47$
baryons have different masses. The $15$ resonances are in good agreement
with experimental data [25]. We have predicted $32$ masses of baryons.

In the framework of the proposed approximate method of solving the
relativistic three-particle problem, we have obtained a satisfactory
spectrum of $N=2$ level baryons. The important problem is the mixing
the states of baryons and the five quark systems (cryptoexotic baryons [31]
or hybrid baryons [32]).

\vskip2.0ex
{\large \bf 6. Conclusion.}
\vskip2.0ex

In the papers [22, 23] the relativistic generalization of Faddeev equations
in the framework of dispersion relation are constructed. We calculated
the $S$-wave baryon masses using the method based on the extraction of
leading singularities of the amplitude. The behavior of electromagnetic
form factor of the nucleon and hyperon in the region of low and
intermediate momentum transfers is determined by [33]. In the framework
of the dispersion relation approach the charge radii of $S$-wave baryon
multiplets with $J^p=\frac{1}{2}^+$ are calculated.

In our paper the relativistic description of three particles
amplitudes of $P$-wave baryons are considered. We take into account
the $u, d, s$-quarks. The mass spectrum of nonstrange and strange states
of multiplet ${\bf (70,1^-)}$ are calculated. We use only four parameters
for the calculation of $30$ baryon masses. We take into account the mass
shift of $u, d, s$ quarks which allows us to obtain the $P$-wave baryon
bound states.

In the present paper the relativistic consideration of three particles
amplitudes of $(70,0^+)$ and $(70,2^+)$ excited baryons are given.
We take into account the $u, d, s$-quarks. We have calculated the $47$
masses of resonances $(70,0^+)$, $(70,2^+)$ with only four parameters.
We take into account the mass shift (similar to [24]) for $u, d, s$ quarks
which allows us to obtain the $N=2$ level excited baryon bound states.
We can see that the masses of these upper multiplets are heavier than
lower, that coincides with the nonrelativistic models [34 -- 37].

The lowest states $\Lambda$ $(1,2)$ $(70,0^+)$ mass is equal to
$m=1.627\, GeV$. The baryon resonances $(70,2^+)$ multiplet heavier than
ones of the $(70,0^+)$ multiplet that is similar to the results of the
papers [15, 16].

\vskip2.0ex
{\large \bf Acknowledgments.}
\vskip2.0ex

The authors would like to thank T. Barnes, S. Capstick, S. Chekanov,
Fl. Stancu for useful discussions. The work was carried with the support
of the Russion Ministry of Education (grant 2.1.1.68.26).

\newpage

{\large \bf Appendix I. The wave functions.}
\vskip2.0ex

We consider, for instance, the wave functions of the upper submultiplets
of decuplet $(10,2)$ $J^p=\frac{5}{2}^+$, octets $(8,2)$
$J^p=\frac{5}{2}^+$, $(8,4)$ $J^p=\frac{7}{2}^+$ and singlet $(1,2)$
$J^p=\frac{5}{2}^+$, which are corresponded to the projection orbital
moment $L_z=2$. For the lower multiplets one must use the corresponding
wave functions. $O(3)$ wave functions possess the mixed symmetry and can
be written as:

$$\varphi_{MA}^{O(3)}=\frac{1}{2}\left(
020-200-101+011
\right),\quad\quad
\varphi_{MS}^{O(3)}=\frac{1}{\sqrt{12}}\left(
020+200-2\cdot 002-101-011+2\cdot 110
\right),\eqno (A1)$$

The $SU(2)$ wave functions have the following form:

$$\varphi_{MA}^{SU(2)}=\frac{1}{\sqrt{2}}\left(
\uparrow \downarrow \uparrow-\downarrow \uparrow \uparrow
\right),\quad\quad
\varphi_{MS}^{SU(2)}=\frac{1}{\sqrt{6}}\left(
\uparrow \downarrow \uparrow+\downarrow \uparrow \uparrow-
2\uparrow \uparrow \downarrow
\right),\eqno (A2)$$

\noindent
The $SU(3)_{f}$ wave functions are different for each particles.

\vskip2.0ex
{\bf Multiplet $(10,2)$.}
\vskip2.0ex

The totally symmetric $SU(6)\times O(3)$ wave function for each
decuplet particle is constructed as:

$$\varphi=\frac{1}{\sqrt{2}}\left(
\varphi_{MA}^{SU(6)}\varphi_{MA}^{O(3)}+
\varphi_{MS}^{SU(6)}\varphi_{MS}^{O(3)}
\right)=\frac{1}{\sqrt{2}}\varphi_{S}^{SU(3)}\left(
\varphi_{MA}^{SU(2)}\varphi_{MA}^{O(3)}+
\varphi_{MS}^{SU(2)}\varphi_{MS}^{O(3)}
\right).\eqno (A3)$$

For the $\Sigma^{+}$-hyperon belonging to the decuplet $SU(3)$ the wave
function is:

$$\varphi_{S}^{SU(3)}=\frac{1}{\sqrt{3}}\left(
usu+suu+uus
\right).\eqno (A4)$$

The totally symmetric $SU(6)\times O(3)$ function of $\Sigma^{+}$ from
multiplet $(10,2)$ is given:

$$\varphi_{\Sigma^{+}(10,2)}=\frac{\sqrt{3}}{18}\left(
2\{u^2\downarrow u\uparrow s\uparrow\}+
\{s^2\downarrow u\uparrow u\uparrow\}-
\{u^2\uparrow u\downarrow s\uparrow\}-
\{u^2\uparrow u\uparrow s\downarrow \}-
\{s^2\uparrow u\uparrow u\downarrow \}-\right.$$
$$\left.-2\{u\downarrow u^1\uparrow s^1\uparrow\}-
\{s\downarrow u^1\uparrow u^1\uparrow\}+
\{u\uparrow u^1\downarrow s^1\uparrow\}+
\{u\uparrow u^1\uparrow s^1\downarrow \}+
\{s\uparrow u^1\uparrow u^1\downarrow \}
\right).\eqno (A5)$$

For the $\Delta^{++}$ of multiplet $(10,2)$ the $SU(6)\times O(3)$ wave
function can be obtained by the replacement $u\leftrightarrow s$:

$$\varphi_{\Delta^{++}(10,2)}=\frac{1}{6}\left(
\{u^2\downarrow u\uparrow u\uparrow\}-\{u^2\uparrow u\uparrow u\downarrow\}
-\{u\downarrow u^1\uparrow u^1\uparrow\}
+\{u\uparrow u^1\uparrow u^1\downarrow\}
\right).\eqno (A6)$$

For the $\Xi$ of decuplet $(10,2)$ the results are similar to $\Sigma$
by the replacement $u\leftrightarrow s$ or $d\leftrightarrow s$. The
$\Omega^{-}$ wave function of $(10,2)$ coinsides with $\Delta$
by the replacement $u\rightarrow s$.

\vskip2.0ex
{\bf Multiplet $(8,2)$.}
\vskip2.0ex

The wave functions of octet $(8,2)$ can be constructed to the following
method:

$$\varphi=\frac{1}{\sqrt{2}}\left(
\varphi_{MA}^{SU(6)}\varphi_{MA}^{O(3)}+
\varphi_{MS}^{SU(6)}\varphi_{MS}^{O(3)}
\right),\eqno (A7)$$

\noindent
where

$$\varphi_{MA}^{SU(6)}=\frac{1}{\sqrt{2}}\left(
\varphi_{MS}^{SU(3)}\varphi_{MA}^{SU(2)}+
\varphi_{MA}^{SU(3)}\varphi_{MS}^{SU(2)}
\right),\eqno (A8)$$

$$\varphi_{MS}^{SU(6)}=\frac{1}{\sqrt{2}}\left(
-\varphi_{MS}^{SU(3)}\varphi_{MS}^{SU(2)}+
\varphi_{MA}^{SU(3)}\varphi_{MA}^{SU(2)}
\right).\eqno (A9)$$

In the case of $\Sigma^{+}$ octet the wave functions $\varphi_{MS}^{SU(3)}$
and $\varphi_{MA}^{SU(3)}$ are:

$$\varphi_{MS}^{SU(3)}=\frac{1}{\sqrt{6}}\left(
usu+suu-2uus\right),\quad
\varphi_{MA}^{SU(3)}=\frac{1}{\sqrt{2}}\left(
usu-suu\right).\eqno (A10)$$

Then the symmetric wave function for $\Sigma^{+}$ of $(8,2)$ have the
following form:

$$\varphi_{\Sigma^{+} (8,2)}=\frac{\sqrt{3}}{18}\left(
2\{u^2\uparrow u\downarrow s\uparrow\}+
\{s^2\downarrow u\uparrow u\uparrow\}-
\{u^2\uparrow u\uparrow s\downarrow\}-
\{u^2\downarrow u\uparrow s\uparrow\}-
\{s^2\uparrow u\uparrow u\downarrow\}-\right.$$

$$\left. -2\{u\uparrow u^1\downarrow s^1\uparrow\}-
\{s\downarrow u^1\uparrow u^1\uparrow\}+
\{u\uparrow u^1\uparrow s^1\downarrow\}+
\{u\downarrow u^1\uparrow s^1\uparrow\}+
\{s\uparrow u^1\uparrow u^1\downarrow\}
\right).\eqno (A11)$$

The nucleon functions of $(8,2)$ can be constructed from $\Sigma^{+}$
by the replacement $s\leftrightarrow d$, and the functions of $\Xi^{0}$
by the replacement $u\leftrightarrow s$.

In the case of the $\Lambda^{0}$ the $SU(3)$ wave functions
$\varphi_{MS}^{SU(3)}$ and $\varphi_{MA}^{SU(3)}$ are:

$$\varphi_{MS}^{SU(3)}=\frac{1}{2}\left(
dsu-usd+sdu-sud
\right),\eqno (A12)$$

$$\varphi_{MA}^{SU(3)}=\frac{\sqrt{3}}{6}\left(
sdu-sud+usd-dsu-2dus+2uds
\right).\eqno (A13)$$

As result, we have obtain the symmetric $SU(6)\times O(3)$ wave function
for $\Lambda^{0}$ of $(8,2)$:

$$\varphi_{\Lambda^{0} (8,2)}=\frac{\sqrt{2}}{12}\left(
\{u^2\uparrow d\uparrow s\downarrow\}-
\{u^2\downarrow d\uparrow s\uparrow\}-
\{d^2\uparrow u\uparrow s\downarrow\}+
\{d^2\downarrow u\uparrow s\uparrow\}-\right.$$

$$\left. -\{s^2\uparrow u\uparrow d\downarrow\}+
\{s^2\uparrow u\downarrow d\uparrow\}-
\{u\uparrow d^1\uparrow s^1\downarrow\}-
\{u\downarrow d^1\uparrow s^1\uparrow\}+\right.$$

$$\left. +\{d\uparrow u^1\uparrow s^1\downarrow\}-
\{d\downarrow u^1\uparrow s^1\uparrow\}+
\{s\uparrow u^1\uparrow d^1\downarrow\}-
\{s\uparrow u^1\downarrow d^1\uparrow\}
\right).\eqno (A14)$$

\vskip2.0ex
{\bf Multiplet $(8,4)$.}
\vskip2.0ex

The wave functions of octet $(8,4)$ are constructed as similar to the cases
of $(10,2)$ and $(8,2)$ multiplets:

$$\varphi=\frac{1}{\sqrt{2}}\left(
\varphi_{MA}^{SU(6)}\varphi_{MA}^{O(3)}+
\varphi_{MS}^{SU(6)}\varphi_{MS}^{O(3)}
\right),\eqno (A15)$$

\noindent
here

$$\varphi_{MA}^{SU(6)}=
\varphi_{MA}^{SU(3)}\varphi_{S}^{SU(2)},\quad
\varphi_{MS}^{SU(6)}=
\varphi_{MS}^{SU(3)}\varphi_{S}^{SU(2)}.\eqno (A16)$$

The $SU(2)$ function is totally symmetric:

$$\varphi_{S}^{SU(2)}=\uparrow\uparrow\uparrow ,\eqno (A17)$$

\noindent
and $\varphi_{MS}^{SU(3)}$, $\varphi_{MA}^{SU(3)}$ similar to $(8,2)$.

For the $\Sigma^{+}$ of $(8,4)$ we can calculate:

$$\varphi_{\Sigma^{+} (8,4)}=\frac{1}{6}\left(
\{s^2\uparrow u\uparrow u\uparrow\}-
\{u^2\uparrow u\uparrow s\uparrow\}-
\{s\uparrow u^1\uparrow u^1\uparrow\}+
\{u\uparrow u^1\uparrow s^1\uparrow\}
\right).\eqno (A18)$$

For the nucleon $N$ of $(8,4)$ the results are similar to $\Sigma^{+}$
of $(8,4)$ by replacement $s\rightarrow d$; and for $\Xi^{0}$ by
replacement $u\leftrightarrow s$.

For $\Lambda^{0}$ of $(8,4)$:

$$\varphi_{\Lambda^{0} (8,4)}=\frac{\sqrt{6}}{12}\left(
-\{u^2\uparrow d\uparrow s\uparrow\}+
\{d^2\uparrow u\uparrow s\uparrow\}+
\{u\uparrow d^1\uparrow s^1\uparrow\}-
\{d\uparrow u^1\uparrow s^1\uparrow\}
\right).\eqno (A19)$$

\vskip2.0ex
{\bf Multiplet $(1,2)$.}
\vskip2.0ex

In the case of $\Lambda^{0}_{1}$ singlet of $(1,2)$ the tolally symmetric
$SU(6)\times O(3)$ function must be constructed in the form:

$$\varphi=\varphi_{A}^{SU(3)}\varphi_{A}^{SU(2)\times O(3)},\eqno (A20)$$

$$\varphi_{A}^{SU(3)}=\frac{1}{\sqrt{6}}\left(
sdu-sud+usd-dsu+dus-uds
\right),\eqno (A21)$$

$$\varphi_{A}^{SU(2)\times O(3)}=\frac{1}{\sqrt{2}}\left(
\varphi_{MS}^{SU(2)}\varphi_{MA}^{O(3)}-
\varphi_{MA}^{SU(2)}\varphi_{MS}^{O(3)}
\right).\eqno (A22)$$

Then, we have calculated the $\varphi_{\Lambda^{0}_{1} (1,2)}$:

$$\varphi_{\Lambda^{0}_{1} (1,2)}=\frac{\sqrt{3}}{6}\left(
-\{u^2\uparrow d\uparrow s\downarrow\}+
\{u^2\uparrow d\downarrow s\uparrow\}+
\{d^2\uparrow u\uparrow s\downarrow\}-
\{d^2\uparrow u\downarrow s\uparrow\}-
\right.$$

$$\left. -\{s^2\uparrow u\uparrow d\downarrow\}+
\{s^2\uparrow u\downarrow d\uparrow\}+
\{u\uparrow d^1\uparrow s^1\downarrow\}-
\{u\uparrow d^1\downarrow s^1\uparrow\}-\right.$$

$$\left. -\{d\uparrow u^1\uparrow s^1\downarrow\}+
\{d\uparrow u^1\downarrow s^1\uparrow\}+
\{s\uparrow u^1\uparrow d^1\downarrow\}-
\{s\uparrow u^1\downarrow d^1\uparrow\}
\right).\eqno (A23)$$

\vskip2.0ex
{\large \bf Appendix II. The system equations of reduced amplitudes of the
multiplets $(70,0^+)$ and $(70,2^+)$.}
\vskip2.0ex
{\bf Multiplet $(10,2)$.}
\vskip2.0ex

$\Delta$ $\frac{5}{2} ^{+}$ $(10,2)$ $(70,2^+)$:

$$
\begin{array}{l}
\alpha_1^{s0+}=\lambda+
\frac{1}{4}\,\,\alpha_1^{s0+}\,\, M_{1^{s+} 1^{s+}}
+\frac{1}{4}\,\,\alpha_1^{p2+}\,\, M_{1^{s+} 3^{d+}}
+\frac{3}{4}\,\,\alpha_0^{d2+}\,\, M_{1^{s+} 2^{d+}}
+\frac{3}{4}\,\,\alpha_0^{p1-}\,\, M_{1^{s+} 1^{p-}}
\hskip9.0ex 1^{s+}\\
\\
\alpha_1^{p2+}=\lambda+
\frac{1}{4}\,\,\alpha_1^{s0+}\,\, M_{3^{d+} 1^{s+}}
+\frac{1}{4}\,\,\alpha_1^{p2+}\,\, M_{3^{d+} 3^{d+}}
+\frac{3}{4}\,\,\alpha_0^{d2+}\,\, M_{3^{d+} 2^{d+}}
+\frac{3}{4}\,\,\alpha_0^{p1-}\,\, M_{3^{d+} 1^{p-}}
\hskip9.0ex 3^{d+}\\
\\
\alpha_0^{d2+}=\lambda+
\frac{3}{4}\,\,\alpha_1^{s0+}\,\, M_{2^{d+} 1^{s+}}
+\frac{3}{4}\,\,\alpha_1^{p2+}\,\, M_{2^{d+} 3^{d+}}
+\frac{1}{4}\,\,\alpha_0^{d2+}\,\, M_{2^{d+} 2^{d+}}
+\frac{1}{4}\,\,\alpha_0^{p1-}\,\, M_{2^{d+} 1^{p-}}
\hskip9.0ex 2^{d+}\\
\\
\alpha_0^{p1-}=\lambda+
\frac{3}{4}\,\,\alpha_1^{s0+}\,\, M_{1^{p-} 1^{s+}}
+\frac{3}{4}\,\,\alpha_1^{p2+}\,\, M_{1^{p-} 3^{d+}}
+\frac{1}{4}\,\,\alpha_0^{d2+}\,\, M_{1^{p-} 2^{d+}}
+\frac{1}{4}\,\,\alpha_0^{p1-}\,\, M_{1^{p-} 1^{p-}}\,\, .
\hskip7.0ex 1^{p-}\\
\end{array}
\eqno (A24)$$

$\Sigma$ $\frac{5}{2} ^{+}$ $(10,2)$ $(70,2^+)$:

$$
\begin{array}{l}
\alpha_1^{s0+}=\lambda+
\frac{1}{4}\,\,\alpha_1^{s0s+}\,\, M_{1^{s+} 1^{s+}_s}
+\frac{1}{4}\,\,\alpha_1^{p2s+}\,\, M_{1^{s+} 3^{d+}_s}
+\frac{3}{4}\,\,\alpha_0^{d2s+}\,\, M_{1^{s+} 2^{d+}_s}
+\frac{3}{4}\,\,\alpha_0^{p1s-}\,\, M_{1^{s+} 1^{p-}_s}
\hskip4.0ex 1^{s+}\\
\\
\alpha_1^{p2+}=\lambda+
\frac{1}{4}\,\,\alpha_1^{s0s+}\,\, M_{3^{d+} 1^{s+}_s}
+\frac{1}{4}\,\,\alpha_1^{p2s+}\,\, M_{3^{d+} 3^{d+}_s}
+\frac{3}{4}\,\,\alpha_0^{d2s+}\,\, M_{3^{d+} 2^{d+}_s}
+\frac{3}{4}\,\,\alpha_0^{p1s-}\,\, M_{3^{d+} 1^{p-}_s}
\hskip4.0ex 3^{d+}\\
\\
\alpha_1^{s0s+}=\lambda+
\frac{1}{2}\,\,\alpha_1^{s0+}\,\, M_{1^{s+}_s 1^{s+}}
+\frac{1}{2}\,\,\alpha_1^{p2+}\,\, M_{1^{s+}_s 3^{d+}}
-\frac{1}{4}\,\,\alpha_1^{s0s+}\,\, M_{1^{s+}_s 1^{s+}_s}
-\frac{1}{4}\,\,\alpha_1^{p2s+}\,\,M_{1^{s+}_s 3^{d+}_s}
\hskip5.5ex 1^{s+}_s\\
\\
\hskip6ex
+\frac{3}{4}\,\,\alpha_0^{d2s+}\,\,M_{1^{s+}_s 2^{d+}_s}
+\frac{3}{4}\,\,\alpha_0^{p1s-}\,\,M_{1^{s+}_s 1^{p-}_s}\\
\\
\alpha_1^{p2s+}=\lambda+
\frac{1}{2}\,\,\alpha_1^{s0+}\,\, M_{3^{d+}_s 1^{s+}}
+\frac{1}{2}\,\,\alpha_1^{p2+}\,\, M_{3^{d+}_s 3^{d+}}
-\frac{1}{4}\,\,\alpha_1^{s0s+}\,\, M_{3^{d+}_s 1^{s+}_s}
-\frac{1}{4}\,\,\alpha_1^{p2s+}\,\,M_{3^{d+}_s 3^{d+}_s}
\hskip5.5ex 3^{d+}_s\\
\\
\hskip6ex
+\frac{3}{4}\,\,\alpha_0^{d2s+}\,\,M_{3^{d+}_s 2^{d+}_s}
+\frac{3}{4}\,\,\alpha_0^{p1s-}\,\,M_{3^{d+}_s 1^{p-}_s}\\
\\
\alpha_0^{d2s+}=\lambda+
\frac{1}{2}\,\,\alpha_1^{s0+}\,\, M_{2^{d+}_s 1^{s+}}
+\frac{1}{2}\,\,\alpha_1^{p2+}\,\, M_{2^{d+}_s 3^{d+}}
+\frac{1}{4}\,\,\alpha_1^{s0s+}\,\, M_{2^{d+}_s 1^{s+}_s}
+\frac{1}{4}\,\,\alpha_1^{p2s+}\,\,M_{2^{d+}_s 3^{d+}_s}
\hskip5.5ex 2^{d+}_s\\
\\
\hskip6ex
+\frac{1}{4}\,\,\alpha_0^{d2s+}\,\,M_{2^{d+}_s 2^{d+}_s}
+\frac{1}{4}\,\,\alpha_0^{p1s-}\,\,M_{2^{d+}_s 1^{p-}_s}\\
\\
\alpha_0^{p1s-}=\lambda+
\frac{1}{2}\,\,\alpha_1^{s0+}\,\, M_{1^{p-}_s 1^{s+}}
+\frac{1}{2}\,\,\alpha_1^{p2+}\,\, M_{1^{p-}_s 3^{d+}}
+\frac{1}{4}\,\,\alpha_1^{s0s+}\,\, M_{1^{p-}_s 1^{s+}_s}
+\frac{1}{4}\,\,\alpha_1^{p2s+}\,\,M_{1^{p-}_s 3^{d+}_s}
\hskip5.0ex 1^{p-}_s\\
\\
\hskip6ex
+\frac{1}{4}\,\,\alpha_0^{d2s+}\,\,M_{1^{p-}_s 2^{d+}_s}
+\frac{1}{4}\,\,\alpha_0^{p1s-}\,\,M_{1^{p-}_s 1^{p-}_s}\,\, .\\
\end{array}
\eqno (A25)$$

The $\Xi$ $\frac{5}{2} ^{+}$ $(10,2)$ $(70,2^+)$ reduced equations are
similar to the $\Sigma$ $\frac{5}{2} ^{+}$ $(10,2)$ $(70,2^+)$ with the
replacement $u\leftrightarrow s$. The $\Omega$ $\frac{5}{2} ^{+}$ $(10,2)$
$(70,2^+)$ reduced equations are constructed by the replacement
$s\leftrightarrow u$ for the $\Delta$ $\frac{5}{2} ^{+}$ $(10,2)$
$(70,2^+)$.

The analogous results are obtained if we have considered the spin
$J^p=\frac{3}{2} ^{+}$ ($(70,2^+)$), $J^p=\frac{1}{2} ^{+}$ ($(70,0^+)$).

\newpage
{\bf Multiplet $(8,2)$.}
\vskip2.0ex

$N$ $\frac{5}{2} ^{+}$ $(8,2)$ $(70,2^+)$:

$$
\begin{array}{l}
\alpha_1^{s0+}=\lambda-
\frac{1}{8}\,\,\alpha_1^{s0+}\,\, M_{1^{s+} 1^{s+}}
-\frac{1}{8}\,\,\alpha_1^{p2+}\,\, M_{1^{s+} 3^{d+}}
+\frac{3}{8}\,\,\alpha_0^{s0+}\,\, M_{1^{s+} 0^{s+}}
+\frac{3}{8}\,\,\alpha_0^{p2+}\,\, M_{1^{s+} 2^{d+}}
\hskip5.5ex 1^{s+}\\
\\
\hskip6ex
+\frac{3}{8}\,\,\alpha_1^{d2+}\,\, M_{1^{s+} 3^{d+}}
+\frac{3}{8}\,\,\alpha_1^{p1-}\,\, M_{1^{s+} 2^{p-}}
+\frac{3}{8}\,\,\alpha_0^{d2+}\,\, M_{1^{s+} 2^{d+}}
+\frac{3}{8}\,\,\alpha_0^{p1-}\,\, M_{1^{s+} 1^{p-}}\\
\\
\alpha_1^{p2+}=\lambda-
\frac{1}{8}\,\,\alpha_1^{s0+}\,\, M_{3^{d+} 1^{s+}}
-\frac{1}{8}\,\,\alpha_1^{p2+}\,\, M_{3^{d+} 3^{d+}}
+\frac{3}{8}\,\,\alpha_0^{s0+}\,\, M_{3^{d+} 0^{s+}}
+\frac{3}{8}\,\,\alpha_0^{p2+}\,\, M_{3^{d+} 2^{d+}}
\hskip5.5ex 3^{d+}\\
\\
\hskip6ex
+\frac{3}{8}\,\,\alpha_1^{d2+}\,\, M_{3^{d+} 3^{d+}}
+\frac{3}{8}\,\,\alpha_1^{p1-}\,\, M_{3^{d+} 2^{p-}}
+\frac{3}{8}\,\,\alpha_0^{d2+}\,\, M_{3^{d+} 2^{d+}}
+\frac{3}{8}\,\,\alpha_0^{p1-}\,\, M_{3^{d+} 1^{p-}}\\
\\
\alpha_0^{s0+}=\lambda+
\frac{3}{8}\,\,\alpha_1^{s0+}\,\, M_{0^{s+} 1^{s+}}
+\frac{3}{8}\,\,\alpha_1^{p2+}\,\, M_{0^{s+} 3^{d+}}
-\frac{1}{8}\,\,\alpha_0^{s0+}\,\, M_{0^{s+} 0^{s+}}
-\frac{1}{8}\,\,\alpha_0^{p2+}\,\, M_{0^{s+} 2^{d+}}
\hskip5.5ex 0^{s+}\\
\\
\hskip6ex
+\frac{3}{8}\,\,\alpha_1^{d2+}\,\, M_{0^{s+} 3^{d+}}
+\frac{3}{8}\,\,\alpha_1^{p1-}\,\, M_{0^{s+} 2^{p-}}
+\frac{3}{8}\,\,\alpha_0^{d2+}\,\, M_{0^{s+} 2^{d+}}
+\frac{3}{8}\,\,\alpha_0^{p1-}\,\, M_{0^{s+} 1^{p-}}\\
\\
\alpha_0^{p2+}=\lambda+
\frac{3}{8}\,\,\alpha_1^{s0+}\,\, M_{2^{d+} 1^{s+}}
+\frac{3}{8}\,\,\alpha_1^{p2+}\,\, M_{2^{d+} 3^{d+}}
-\frac{1}{8}\,\,\alpha_0^{s0+}\,\, M_{2^{d+} 0^{s+}}
-\frac{1}{8}\,\,\alpha_0^{p2+}\,\, M_{2^{d+} 2^{d+}}
\hskip5.5ex 2^{d+}\\
\\
\hskip6ex
+\frac{3}{8}\,\,\alpha_1^{d2+}\,\, M_{2^{d+} 3^{d+}}
+\frac{3}{8}\,\,\alpha_1^{p1-}\,\, M_{2^{d+} 2^{p-}}
+\frac{3}{8}\,\,\alpha_0^{d2+}\,\, M_{2^{d+} 2^{d+}}
+\frac{3}{8}\,\,\alpha_0^{p1-}\,\, M_{2^{d+} 1^{p-}}\\
\\
\alpha_1^{d2+}=\lambda+
\frac{3}{8}\,\,\alpha_1^{s0+}\,\, M_{3^{d+} 1^{s+}}
+\frac{3}{8}\,\,\alpha_1^{p2+}\,\, M_{3^{d+} 3^{d+}}
+\frac{3}{8}\,\,\alpha_0^{s0+}\,\, M_{3^{d+} 0^{s+}}
+\frac{3}{8}\,\,\alpha_0^{p2+}\,\, M_{3^{d+} 2^{d+}}
\hskip5.5ex 3^{d+}\\
\\
\hskip6ex
-\frac{1}{8}\,\,\alpha_1^{d2+}\,\, M_{3^{d+} 3^{d+}}
-\frac{1}{8}\,\,\alpha_1^{p1-}\,\, M_{3^{d+} 2^{p-}}
+\frac{3}{8}\,\,\alpha_0^{d2+}\,\, M_{3^{d+} 2^{d+}}
+\frac{3}{8}\,\,\alpha_0^{p1-}\,\, M_{3^{d+} 1^{p-}}\\
\\
\alpha_1^{p1-}=\lambda+
\frac{3}{8}\,\,\alpha_1^{s0+}\,\, M_{2^{p-} 1^{s+}}
+\frac{3}{8}\,\,\alpha_1^{p2+}\,\, M_{2^{p-} 3^{d+}}
+\frac{3}{8}\,\,\alpha_0^{s0+}\,\, M_{2^{p-} 0^{s+}}
+\frac{3}{8}\,\,\alpha_0^{p2+}\,\, M_{2^{p-} 2^{d+}}
\hskip5.5ex 2^{p-}\\
\\
\hskip6ex
-\frac{1}{8}\,\,\alpha_1^{d2+}\,\, M_{2^{p-} 3^{d+}}
-\frac{1}{8}\,\,\alpha_1^{p1-}\,\, M_{2^{p-} 2^{p-}}
+\frac{3}{8}\,\,\alpha_0^{d2+}\,\, M_{2^{p-} 2^{d+}}
+\frac{3}{8}\,\,\alpha_0^{p1-}\,\, M_{2^{p-} 1^{p-}}\\
\\
\alpha_0^{d2+}=\lambda+
\frac{3}{8}\,\,\alpha_1^{s0+}\,\, M_{2^{d+} 1^{s+}}
+\frac{3}{8}\,\,\alpha_1^{p2+}\,\, M_{2^{d+} 3^{d+}}
+\frac{3}{8}\,\,\alpha_0^{s0+}\,\, M_{2^{d+} 0^{s+}}
+\frac{3}{8}\,\,\alpha_0^{p2+}\,\, M_{2^{d+} 2^{d+}}
\hskip5.5ex 2^{d+}\\
\\
\hskip6ex
+\frac{3}{8}\,\,\alpha_1^{d2+}\,\, M_{2^{d+} 3^{d+}}
+\frac{3}{8}\,\,\alpha_1^{p1-}\,\, M_{2^{d+} 2^{p-}}
-\frac{1}{8}\,\,\alpha_0^{d2+}\,\, M_{2^{d+} 2^{d+}}
-\frac{1}{8}\,\,\alpha_0^{p1-}\,\, M_{2^{d+} 1^{p-}}\\
\\
\alpha_0^{p1-}=\lambda+
\frac{3}{8}\,\,\alpha_1^{s0+}\,\, M_{1^{p-} 1^{s+}}
+\frac{3}{8}\,\,\alpha_1^{p2+}\,\, M_{1^{p-} 3^{d+}}
+\frac{3}{8}\,\,\alpha_0^{s0+}\,\, M_{1^{p-} 0^{s+}}
+\frac{3}{8}\,\,\alpha_0^{p2+}\,\, M_{1^{p-} 2^{d+}}
\hskip5.5ex 1^{p-}\\
\\
\hskip6ex
+\frac{3}{8}\,\,\alpha_1^{d2+}\,\, M_{1^{p-} 3^{d+}}
+\frac{3}{8}\,\,\alpha_1^{p1-}\,\, M_{1^{p-} 2^{p-}}
-\frac{1}{8}\,\,\alpha_0^{d2+}\,\, M_{1^{p-} 2^{d+}}
-\frac{1}{8}\,\,\alpha_0^{p1-}\,\, M_{1^{p-} 1^{p-}}\,\, .\\
\end{array}
\eqno (A26)$$

$\Sigma$ $\frac{5}{2} ^{+}$ $(8,2)$ $(70,2^+)$:

$$
\begin{array}{l}
\alpha_1^{s0+}=\lambda-
\frac{1}{8}\,\,\alpha_1^{s0s+}\,\, M_{1^{s+} 1^{s+}_s}
-\frac{1}{8}\,\,\alpha_1^{p2s+}\,\, M_{1^{s+} 3^{d+}_s}
+\frac{3}{8}\,\,\alpha_0^{s0s+}\,\, M_{1^{s+} 0^{s+}_s}
+\frac{3}{8}\,\,\alpha_0^{p2s+}\,\, M_{1^{s+} 2^{d+}_s}
\hskip4.0ex 1^{s+}\\
\\
\hskip5.5ex
+\frac{3}{8}\,\,\alpha_1^{d2s+}\,\, M_{1^{s+} 3^{d+}_s}
+\frac{3}{8}\,\,\alpha_1^{p1s-}\,\, M_{1^{s+} 2^{p-}_s}
+\frac{3}{8}\,\,\alpha_0^{d2s+}\,\, M_{1^{s+} 2^{d+}_s}
+\frac{3}{8}\,\,\alpha_0^{p1s-}\,\, M_{1^{s+} 1^{p-}_s}\\
\\
\alpha_1^{p2+}=\lambda-
\frac{1}{8}\,\,\alpha_1^{s0s+}\,\, M_{3^{d+} 1^{s+}_s}
-\frac{1}{8}\,\,\alpha_1^{p2s+}\,\, M_{3^{d+} 3^{d+}_s}
+\frac{3}{8}\,\,\alpha_0^{s0s+}\,\, M_{3^{d+} 0^{s+}_s}
+\frac{3}{8}\,\,\alpha_0^{p2s+}\,\, M_{3^{d+} 2^{d+}_s}
\hskip4.0ex 3^{d+}\\
\\
\hskip5.5ex
+\frac{3}{8}\,\,\alpha_1^{d2s+}\,\, M_{3^{d+} 3^{d+}_s}
+\frac{3}{8}\,\,\alpha_1^{p1s-}\,\, M_{3^{d+} 2^{p-}_s}
+\frac{3}{8}\,\,\alpha_0^{d2s+}\,\, M_{3^{d+} 2^{d+}_s}
+\frac{3}{8}\,\,\alpha_0^{p1s-}\,\, M_{3^{d+} 1^{p-}_s}\\
\\
\alpha_1^{s0s+}=\lambda+
\frac{1}{2}\,\,\alpha_1^{s0+}\,\, M_{1^{s+}_s 1^{s+}}
+\frac{1}{2}\,\,\alpha_1^{p2+}\,\, M_{1^{s+}_s 3^{d+}}
-\frac{5}{8}\,\,\alpha_1^{s0s+}\,\, M_{1^{s+}_s 1^{s+}_s}
-\frac{5}{8}\,\,\alpha_1^{p2s+}\,\, M_{1^{s+}_s 3^{d+}_s}
\hskip5.5ex 1^{s+}_s\\
\\
\hskip6ex
+\frac{3}{8}\,\,\alpha_0^{s0s+}\,\, M_{1^{s+}_s 0^{s+}_s}
+\frac{3}{8}\,\,\alpha_0^{p2s+}\,\, M_{1^{s+}_s 2^{d+}_s}
+\frac{3}{8}\,\,\alpha_1^{d2s+}\,\, M_{1^{s+}_s 3^{d+}_s}
+\frac{3}{8}\,\,\alpha_1^{p1s-}\,\, M_{1^{s+}_s 2^{p-}_s}\\
\\
\hskip6ex
+\frac{3}{8}\,\,\alpha_0^{d2s+}\,\, M_{1^{s+}_s 2^{d+}_s}
+\frac{3}{8}\,\,\alpha_0^{p1s-}\,\, M_{1^{s+}_s 1^{p-}_s}\\
\end{array}
$$

\newpage

$$
\begin{array}{l}
\alpha_1^{p2s+}=\lambda+
\frac{1}{2}\,\,\alpha_1^{s0+}\,\, M_{3^{d+}_s 1^{s+}}
+\frac{1}{2}\,\,\alpha_1^{p2+}\,\, M_{3^{d+}_s 3^{d+}}
-\frac{5}{8}\,\,\alpha_1^{s0s+}\,\, M_{3^{d+}_s 1^{s+}_s}
-\frac{5}{8}\,\,\alpha_1^{p2s+}\,\, M_{3^{d+}_s 3^{d+}_s}
\hskip5.5ex 3^{d+}_s\\
\\
\hskip6ex
+\frac{3}{8}\,\,\alpha_0^{s0s+}\,\, M_{3^{d+}_s 0^{s+}_s}
+\frac{3}{8}\,\,\alpha_0^{p2s+}\,\, M_{3^{d+}_s 2^{d+}_s}
+\frac{3}{8}\,\,\alpha_1^{d2s+}\,\, M_{3^{d+}_s 3^{d+}_s}
+\frac{3}{8}\,\,\alpha_1^{p1s-}\,\, M_{3^{d+}_s 2^{p-}_s}\\
\\
\hskip6ex
+\frac{3}{8}\,\,\alpha_0^{d2s+}\,\, M_{3^{d+}_s 2^{d+}_s}
+\frac{3}{8}\,\,\alpha_0^{p1s-}\,\, M_{3^{d+}_s 1^{p-}_s}\\
\\
\alpha_0^{s0s+}=\lambda+
\frac{1}{2}\,\,\alpha_1^{s0+}\,\, M_{0^{s+}_s 1^{s+}}
+\frac{1}{2}\,\,\alpha_1^{p2+}\,\, M_{0^{s+}_s 3^{d+}}
-\frac{1}{8}\,\,\alpha_1^{s0s+}\,\, M_{0^{s+}_s 1^{s+}_s}
-\frac{1}{8}\,\,\alpha_1^{p2s+}\,\, M_{0^{s+}_s 3^{d+}_s}
\hskip5.5ex 0^{s+}_s\\
\\
\hskip6ex
-\frac{1}{8}\,\,\alpha_0^{s0s+}\,\, M_{0^{s+}_s 0^{s+}_s}
-\frac{1}{8}\,\,\alpha_0^{p2s+}\,\, M_{0^{s+}_s 2^{d+}_s}
+\frac{3}{8}\,\,\alpha_1^{d2s+}\,\, M_{0^{s+}_s 3^{d+}_s}
+\frac{3}{8}\,\,\alpha_1^{p1s-}\,\, M_{0^{s+}_s 2^{p-}_s}\\
\\
\hskip6ex
+\frac{3}{8}\,\,\alpha_0^{d2s+}\,\, M_{0^{s+}_s 2^{d+}_s}
+\frac{3}{8}\,\,\alpha_0^{p1s-}\,\, M_{0^{s+}_s 1^{p-}_s}\\
\\
\alpha_0^{p2s+}=\lambda+
\frac{1}{2}\,\,\alpha_1^{s0+}\,\, M_{2^{d+}_s 1^{s+}}
+\frac{1}{2}\,\,\alpha_1^{p2+}\,\, M_{2^{d+}_s 3^{d+}}
-\frac{1}{8}\,\,\alpha_1^{s0s+}\,\, M_{2^{d+}_s 1^{s+}_s}
-\frac{1}{8}\,\,\alpha_1^{p2s+}\,\, M_{2^{d+}_s 3^{d+}_s}
\hskip5.5ex 2^{d+}_s\\
\\
\hskip6ex
-\frac{1}{8}\,\,\alpha_0^{s0s+}\,\, M_{2^{d+}_s 0^{s+}_s}
-\frac{1}{8}\,\,\alpha_0^{p2s+}\,\, M_{2^{d+}_s 2^{d+}_s}
+\frac{3}{8}\,\,\alpha_1^{d2s+}\,\, M_{2^{d+}_s 3^{d+}_s}
+\frac{3}{8}\,\,\alpha_1^{p1s-}\,\, M_{2^{d+}_s 2^{p-}_s}\\
\\
\hskip6ex
+\frac{3}{8}\,\,\alpha_0^{d2s+}\,\, M_{2^{d+}_s 2^{d+}_s}
+\frac{3}{8}\,\,\alpha_0^{p1s-}\,\, M_{2^{d+}_s 1^{p-}_s}\\
\\
\alpha_1^{d2s+}=\lambda+
\frac{1}{2}\,\,\alpha_1^{s0+}\,\, M_{3^{d+}_s 1^{s+}}
+\frac{1}{2}\,\,\alpha_1^{p2+}\,\, M_{3^{d+}_s 3^{d+}}
-\frac{1}{8}\,\,\alpha_1^{s0s+}\,\, M_{3^{d+}_s 1^{s+}_s}
-\frac{1}{8}\,\,\alpha_1^{p2s+}\,\, M_{3^{d+}_s 3^{d+}_s}
\hskip5.5ex 3^{d+}_s\\
\\
\hskip6ex
+\frac{3}{8}\,\,\alpha_0^{s0s+}\,\, M_{3^{d+}_s 0^{s+}_s}
+\frac{3}{8}\,\,\alpha_0^{p2s+}\,\, M_{3^{d+}_s 2^{d+}_s}
-\frac{1}{8}\,\,\alpha_1^{d2s+}\,\, M_{3^{d+}_s 3^{d+}_s}
-\frac{1}{8}\,\,\alpha_1^{p1s-}\,\, M_{3^{d+}_s 2^{p-}_s}\\
\\
\hskip6ex
+\frac{3}{8}\,\,\alpha_0^{d2s+}\,\, M_{3^{d+}_s 2^{d+}_s}
+\frac{3}{8}\,\,\alpha_0^{p1s-}\,\, M_{3^{d+}_s 1^{p-}_s}\\
\\
\alpha_1^{p1s-}=\lambda+
\frac{1}{2}\,\,\alpha_1^{s0+}\,\, M_{2^{p-}_s 1^{s+}}
+\frac{1}{2}\,\,\alpha_1^{p2+}\,\, M_{2^{p-}_s 3^{d+}}
-\frac{1}{8}\,\,\alpha_1^{s0s+}\,\, M_{2^{p-}_s 1^{s+}_s}
-\frac{1}{8}\,\,\alpha_1^{p2s+}\,\, M_{2^{p-}_s 3^{d+}_s}
\hskip5.5ex 2^{p-}_s\\
\\
\hskip6ex
+\frac{3}{8}\,\,\alpha_0^{s0s+}\,\, M_{2^{p-}_s 0^{s+}_s}
+\frac{3}{8}\,\,\alpha_0^{p2s+}\,\, M_{2^{p-}_s 2^{d+}_s}
-\frac{1}{8}\,\,\alpha_1^{d2s+}\,\, M_{2^{p-}_s 3^{d+}_s}
-\frac{1}{8}\,\,\alpha_1^{p1s-}\,\, M_{2^{p-}_s 2^{p-}_s}\\
\\
\hskip6ex
+\frac{3}{8}\,\,\alpha_0^{d2s+}\,\, M_{2^{p-}_s 2^{d+}_s}
+\frac{3}{8}\,\,\alpha_0^{p1s-}\,\, M_{2^{p-}_s 1^{p-}_s}\\
\\
\alpha_0^{d2s+}=\lambda+
\frac{1}{2}\,\,\alpha_1^{s0+}\,\, M_{2^{d+}_s 1^{s+}}
+\frac{1}{2}\,\,\alpha_1^{p2+}\,\, M_{2^{d+}_s 3^{d+}}
-\frac{1}{8}\,\,\alpha_1^{s0s+}\,\, M_{2^{d+}_s 1^{s+}_s}
-\frac{1}{8}\,\,\alpha_1^{p2s+}\,\, M_{2^{d+}_s 3^{d+}_s}
\hskip5.5ex 2^{d+}_s\\
\\
\hskip6ex
+\frac{3}{8}\,\,\alpha_0^{s0s+}\,\, M_{2^{d+}_s 0^{s+}_s}
+\frac{3}{8}\,\,\alpha_0^{p2s+}\,\, M_{2^{d+}_s 2^{d+}_s}
+\frac{3}{8}\,\,\alpha_1^{d2s+}\,\, M_{2^{d+}_s 3^{d+}_s}
+\frac{3}{8}\,\,\alpha_1^{p1s-}\,\, M_{2^{d+}_s 2^{p-}_s}\\
\\
\hskip6ex
-\frac{1}{8}\,\,\alpha_0^{d2s+}\,\, M_{2^{d+}_s 2^{d+}_s}
-\frac{1}{8}\,\,\alpha_0^{p1s-}\,\, M_{2^{d+}_s 1^{p-}_s}\\
\\
\alpha_0^{p1s-}=\lambda+
\frac{1}{2}\,\,\alpha_1^{s0+}\,\, M_{1^{p-}_s 1^{s+}}
+\frac{1}{2}\,\,\alpha_1^{p2+}\,\, M_{1^{p-}_s 3^{d+}}
-\frac{1}{8}\,\,\alpha_1^{s0s+}\,\, M_{1^{p-}_s 1^{s+}_s}
-\frac{1}{8}\,\,\alpha_1^{p2s+}\,\, M_{1^{p-}_s 3^{d+}_s}
\hskip5.5ex 1^{p-}_s\\
\\
\hskip6ex
+\frac{3}{8}\,\,\alpha_0^{s0s+}\,\, M_{1^{p-}_s 0^{s+}_s}
+\frac{3}{8}\,\,\alpha_0^{p2s+}\,\, M_{1^{p-}_s 2^{d+}_s}
+\frac{3}{8}\,\,\alpha_1^{d2s+}\,\, M_{1^{p-}_s 3^{d+}_s}
+\frac{3}{8}\,\,\alpha_1^{p1s-}\,\, M_{1^{p-}_s 2^{p-}_s}\\
\\
\hskip6ex
-\frac{1}{8}\,\,\alpha_0^{d2s+}\,\, M_{1^{p-}_s 2^{d+}_s}
-\frac{1}{8}\,\,\alpha_0^{p1s-}\,\, M_{1^{p-}_s 1^{p-}_s}\,\, .\\
\end{array}
\eqno (A27)$$

\newpage

$\Lambda$ $\frac{5}{2} ^{+}$ $(8,2)$ $(70,2^+)$:

$$
\begin{array}{l}
\alpha_1^{s0+}=\lambda+
\frac{3}{8}\,\,\alpha_1^{s0s+}\,\, M_{1^{s+} 1^{s+}_s}
+\frac{3}{8}\,\,\alpha_1^{p2s+}\,\, M_{1^{s+} 3^{d+}_s}
+\frac{3}{8}\,\,\alpha_0^{s0s+}\,\, M_{1^{s+} 0^{s+}_s}
+\frac{3}{8}\,\,\alpha_0^{p2s+}\,\, M_{1^{s+} 2^{d+}_s}
\hskip4.0ex 1^{s+}\\
\\
\hskip5.5ex
+\frac{3}{8}\,\,\alpha_1^{d2s+}\,\, M_{1^{s+} 3^{d+}_s}
+\frac{3}{8}\,\,\alpha_1^{p1s-}\,\, M_{1^{s+} 2^{p-}_s}
-\frac{1}{8}\,\,\alpha_0^{d2s+}\,\, M_{1^{s+} 2^{d+}_s}
-\frac{1}{8}\,\,\alpha_0^{p1s-}\,\, M_{1^{s+} 1^{p-}_s}\\
\\
\alpha_1^{p2+}=\lambda+
\frac{3}{8}\,\,\alpha_1^{s0s+}\,\, M_{3^{d+} 1^{s+}_s}
+\frac{3}{8}\,\,\alpha_1^{p2s+}\,\, M_{3^{d+} 3^{d+}_s}
+\frac{3}{8}\,\,\alpha_0^{s0s+}\,\, M_{3^{d+} 0^{s+}_s}
+\frac{3}{8}\,\,\alpha_0^{p2s+}\,\, M_{3^{d+} 2^{d+}_s}
\hskip4.0ex 3^{d+}\\
\\
\hskip5.5ex
+\frac{3}{8}\,\,\alpha_1^{d2s+}\,\, M_{3^{d+} 3^{d+}_s}
+\frac{3}{8}\,\,\alpha_1^{p1s-}\,\, M_{3^{d+} 2^{p-}_s}
-\frac{1}{8}\,\,\alpha_0^{d2s+}\,\, M_{3^{d+} 2^{d+}_s}
-\frac{1}{8}\,\,\alpha_0^{p1s-}\,\, M_{3^{d+} 1^{p-}_s}\\
\\
\alpha_1^{s0s+}=\lambda+
\frac{1}{2}\,\,\alpha_1^{s0+}\,\, M_{1^{s+}_s 1^{s+}}
+\frac{1}{2}\,\,\alpha_1^{p2+}\,\, M_{1^{s+}_s 3^{d+}}
-\frac{1}{8}\,\,\alpha_1^{s0s+}\,\, M_{1^{s+}_s 1^{s+}_s}
-\frac{1}{8}\,\,\alpha_1^{p2s+}\,\, M_{1^{s+}_s 3^{d+}_s}
\hskip5.5ex 1^{s+}_s\\
\\
\hskip6ex
+\frac{3}{8}\,\,\alpha_0^{s0s+}\,\, M_{1^{s+}_s 0^{s+}_s}
+\frac{3}{8}\,\,\alpha_0^{p2s+}\,\, M_{1^{s+}_s 2^{d+}_s}
+\frac{3}{8}\,\,\alpha_1^{d2s+}\,\, M_{1^{s+}_s 3^{d+}_s}
+\frac{3}{8}\,\,\alpha_1^{p1s-}\,\, M_{1^{s+}_s 2^{p-}_s}\\
\\
\hskip6ex
-\frac{1}{8}\,\,\alpha_0^{d2s+}\,\, M_{1^{s+}_s 2^{d+}_s}
-\frac{1}{8}\,\,\alpha_0^{p1s-}\,\, M_{1^{s+}_s 1^{p-}_s}\\
\\
\alpha_1^{p2s+}=\lambda+
\frac{1}{2}\,\,\alpha_1^{s0+}\,\, M_{3^{d+}_s 1^{s+}}
+\frac{1}{2}\,\,\alpha_1^{p2+}\,\, M_{3^{d+}_s 3^{d+}}
-\frac{1}{8}\,\,\alpha_1^{s0s+}\,\, M_{3^{d+}_s 1^{s+}_s}
-\frac{1}{8}\,\,\alpha_1^{p2s+}\,\, M_{3^{d+}_s 3^{d+}_s}
\hskip5.5ex 3^{d+}_s\\
\\
\hskip6ex
+\frac{3}{8}\,\,\alpha_0^{s0s+}\,\, M_{3^{d+}_s 0^{s+}_s}
+\frac{3}{8}\,\,\alpha_0^{p2s+}\,\, M_{3^{d+}_s 2^{d+}_s}
+\frac{3}{8}\,\,\alpha_1^{d2s+}\,\, M_{3^{d+}_s 3^{d+}_s}
+\frac{3}{8}\,\,\alpha_1^{p1s-}\,\, M_{3^{d+}_s 2^{p-}_s}\\
\\
\hskip6ex
-\frac{1}{8}\,\,\alpha_0^{d2s+}\,\, M_{3^{d+}_s 2^{d+}_s}
-\frac{1}{8}\,\,\alpha_0^{p1s-}\,\, M_{3^{d+}_s 1^{p-}_s}\\
\\
\alpha_0^{s0s+}=\lambda+
\frac{1}{2}\,\,\alpha_1^{s0+}\,\, M_{0^{s+}_s 1^{s+}}
+\frac{1}{2}\,\,\alpha_1^{p2+}\,\, M_{0^{s+}_s 3^{d+}}
+\frac{3}{8}\,\,\alpha_1^{s0s+}\,\, M_{0^{s+}_s 1^{s+}_s}
+\frac{3}{8}\,\,\alpha_1^{p2s+}\,\, M_{0^{s+}_s 3^{d+}_s}
\hskip5.5ex 0^{s+}_s\\
\\
\hskip6ex
-\frac{1}{8}\,\,\alpha_0^{s0s+}\,\, M_{0^{s+}_s 0^{s+}_s}
-\frac{1}{8}\,\,\alpha_0^{p2s+}\,\, M_{0^{s+}_s 2^{d+}_s}
+\frac{3}{8}\,\,\alpha_1^{d2s+}\,\, M_{0^{s+}_s 3^{d+}_s}
+\frac{3}{8}\,\,\alpha_1^{p1s-}\,\, M_{0^{s+}_s 2^{p-}_s}\\
\\
\hskip6ex
-\frac{1}{8}\,\,\alpha_0^{d2s+}\,\, M_{0^{s+}_s 2^{d+}_s}
-\frac{1}{8}\,\,\alpha_0^{p1s}-\,\, M_{0^{s+}_s 1^{p-}_s}\\
\\
\alpha_0^{p2s+}=\lambda+
\frac{1}{2}\,\,\alpha_1^{s0+}\,\, M_{2^{d+}_s 1^{s+}}
+\frac{1}{2}\,\,\alpha_1^{p2+}\,\, M_{2^{d+}_s 3^{d+}}
+\frac{3}{8}\,\,\alpha_1^{s0s+}\,\, M_{2^{d+}_s 1^{s+}_s}
+\frac{3}{8}\,\,\alpha_1^{p2s+}\,\, M_{2^{d+}_s 3^{d+}_s}
\hskip5.5ex 2^{d+}_s\\
\\
\hskip6ex
-\frac{1}{8}\,\,\alpha_0^{s0s+}\,\, M_{2^{d+}_s 0^{s+}_s}
-\frac{1}{8}\,\,\alpha_0^{p2s+}\,\, M_{2^{d+}_s 2^{d+}_s}
+\frac{3}{8}\,\,\alpha_1^{d2s+}\,\, M_{2^{d+}_s 3^{d+}_s}
+\frac{3}{8}\,\,\alpha_1^{p1s-}\,\, M_{2^{d+}_s 2^{p-}_s}\\
\\
\hskip6ex
-\frac{1}{8}\,\,\alpha_0^{d2s+}\,\, M_{2^{d+}_s 2^{d+}_s}
-\frac{1}{8}\,\,\alpha_0^{p1s-}\,\, M_{2^{d+}_s 1^{p-}_s}\\
\\
\alpha_1^{d2s+}=\lambda+
\frac{1}{2}\,\,\alpha_1^{s0+}\,\, M_{3^{d+}_s 1^{s+}}
+\frac{1}{2}\,\,\alpha_1^{p2+}\,\, M_{3^{d+}_s 3^{d+}}
+\frac{3}{8}\,\,\alpha_1^{s0s+}\,\, M_{3^{d+}_s 1^{s+}_s}
+\frac{3}{8}\,\,\alpha_1^{p2s+}\,\, M_{3^{d+}_s 3^{d+}_s}
\hskip5.5ex 3^{d+}_s\\
\\
\hskip6ex
+\frac{3}{8}\,\,\alpha_0^{s0s+}\,\, M_{3^{d+}_s 0^{s+}_s}
+\frac{3}{8}\,\,\alpha_0^{p2s+}\,\, M_{3^{d+}_s 2^{d+}_s}
-\frac{1}{8}\,\,\alpha_1^{d2s+}\,\, M_{3^{d+}_s 3^{d+}_s}
-\frac{1}{8}\,\,\alpha_1^{p1s-}\,\, M_{3^{d+}_s 2^{p-}_s}\\
\\
\hskip6ex
-\frac{1}{8}\,\,\alpha_0^{d2s+}\,\, M_{3^{d+}_s 2^{d+}_s}
-\frac{1}{8}\,\,\alpha_0^{p1s-}\,\, M_{3^{d+}_s 1^{p-}_s}\\
\\
\alpha_1^{p1s-}=\lambda+
\frac{1}{2}\,\,\alpha_1^{s0+}\,\, M_{2^{p-}_s 1^{s+}}
+\frac{1}{2}\,\,\alpha_1^{p2+}\,\, M_{2^{p-}_s 3^{d+}}
+\frac{3}{8}\,\,\alpha_1^{s0s+}\,\, M_{2^{p-}_s 1^{s+}_s}
+\frac{3}{8}\,\,\alpha_1^{p2s+}\,\, M_{2^{p-}_s 3^{d+}_s}
\hskip5.5ex 2^{p-}_s\\
\\
\hskip6ex
+\frac{3}{8}\,\,\alpha_0^{s0s+}\,\, M_{2^{p-}_s 0^{s+}_s}
+\frac{3}{8}\,\,\alpha_0^{p2s+}\,\, M_{2^{p-}_s 2^{d+}_s}
-\frac{1}{8}\,\,\alpha_1^{d2s+}\,\, M_{2^{p-}_s 3^{d+}_s}
-\frac{1}{8}\,\,\alpha_1^{p1s-}\,\, M_{2^{p-}_s 2^{p-}_s}\\
\\
\hskip6ex
-\frac{1}{8}\,\,\alpha_0^{d2s+}\,\, M_{2^{p-}_s 2^{d+}_s}
-\frac{1}{8}\,\,\alpha_0^{p1s-}\,\, M_{2^{p-}_s 1^{p-}_s}\\
\end{array}
$$

\newpage

$$
\begin{array}{l}
\alpha_0^{d2s+}=\lambda+
\frac{1}{2}\,\,\alpha_1^{s0+}\,\, M_{2^{d+}_s 1^{s+}}
+\frac{1}{2}\,\,\alpha_1^{p2+}\,\, M_{2^{d+}_s 3^{d+}}
+\frac{3}{8}\,\,\alpha_1^{s0s+}\,\, M_{2^{d+}_s 1^{s+}_s}
+\frac{3}{8}\,\,\alpha_1^{p2s+}\,\, M_{2^{d+}_s 3^{d+}_s}
\hskip5.5ex 2^{d+}_s\\
\\
\hskip6ex
+\frac{3}{8}\,\,\alpha_0^{s0s+}\,\, M_{2^{d+}_s 0^{s+}_s}
+\frac{3}{8}\,\,\alpha_0^{p2s+}\,\, M_{2^{d+}_s 2^{d+}_s}
+\frac{3}{8}\,\,\alpha_1^{d2s+}\,\, M_{2^{d+}_s 3^{d+}_s}
+\frac{3}{8}\,\,\alpha_1^{p1s-}\,\, M_{2^{d+}_s 2^{p-}_s}\\
\\
\hskip6ex
-\frac{5}{8}\,\,\alpha_0^{d2s+}\,\, M_{2^{d+}_s 2^{d+}_s}
-\frac{5}{8}\,\,\alpha_0^{p1s-}\,\, M_{2^{d+}_s 1^{p-}_s}\\
\\
\alpha_0^{p1s-}=\lambda+
\frac{1}{2}\,\,\alpha_1^{s0+}\,\, M_{1^{p-}_s 1^{s+}}
+\frac{1}{2}\,\,\alpha_1^{p2+}\,\, M_{1^{p-}_s 3^{d+}}
+\frac{3}{8}\,\,\alpha_1^{s0s+}\,\, M_{1^{p-}_s 1^{s+}_s}
+\frac{3}{8}\,\,\alpha_1^{p2s+}\,\, M_{1^{p-}_s 3^{d+}_s}
\hskip5.5ex 1^{p-}_s\\
\\
\hskip6ex
+\frac{3}{8}\,\,\alpha_0^{s0s+}\,\, M_{1^{p-}_s 0^{s+}_s}
+\frac{3}{8}\,\,\alpha_0^{p2s+}\,\, M_{1^{p-}_s 2^{d+}_s}
+\frac{3}{8}\,\,\alpha_1^{d2s+}\,\, M_{1^{p-}_s 3^{d+}_s}
+\frac{3}{8}\,\,\alpha_1^{p1s-}\,\, M_{1^{p-}_s 2^{p-}_s}\\
\\
\hskip6ex
-\frac{5}{8}\,\,\alpha_0^{d2s+}\,\, M_{1^{p-}_s 2^{d+}_s}
-\frac{5}{8}\,\,\alpha_0^{p1s-}\,\, M_{1^{p-}_s 1^{p-}_s}\,\, .\\
\end{array}
\eqno (A28)$$

\vskip2.0ex
{\bf Multiplet $(8,4)$.}
\vskip2.0ex

$N$ $\frac{7}{2} ^{+}$ $(8,4)$ $(70,2^+)$:

$$
\begin{array}{l}
\alpha_1^{s0+}=\lambda+
\frac{1}{4}\,\,\alpha_1^{s0+}\,\, M_{1^{s+} 1^{s+}}
+\frac{1}{4}\,\,\alpha_1^{p2+}\,\, M_{1^{s+} 3^{d+}}
+\frac{3}{4}\,\,\alpha_1^{d2+}\,\, M_{1^{s+} 3^{d+}}
+\frac{3}{4}\,\,\alpha_1^{p1-}\,\, M_{1^{s+} 2^{p-}}
\hskip12.0ex 1^{s+}\\
\\
\alpha_1^{p2+}=\lambda+
\frac{1}{4}\,\,\alpha_1^{s0+}\,\, M_{3^{d+} 1^{s+}}
+\frac{1}{4}\,\,\alpha_1^{p2+}\,\, M_{3^{d+} 3^{d+}}
+\frac{3}{4}\,\,\alpha_1^{d2+}\,\, M_{3^{d+} 3^{d+}}
+\frac{3}{4}\,\,\alpha_1^{p1-}\,\, M_{3^{d+} 2^{p-}}
\hskip12.0ex 3^{d+}\\
\\
\alpha_1^{d2+}=\lambda+
\frac{3}{4}\,\,\alpha_1^{s0+}\,\, M_{3^{d+} 1^{s+}}
+\frac{3}{4}\,\,\alpha_1^{p2+}\,\,M_{3^{d+} 3^{d+}}
+\frac{1}{4}\,\,\alpha_1^{d2+}\,\,M_{3^{d+} 3^{d+}}
+\frac{1}{4}\,\,\alpha_1^{p1-}\,\,M_{3^{d+} 2^{p-}}
\hskip12ex 3^{d+}\\
\\
\alpha_1^{p1-}=\lambda+
\frac{3}{4}\,\,\alpha_1^{s0+}\,\, M_{2^{p-} 1^{s+}}
+\frac{3}{4}\,\,\alpha_1^{p2+}\,\,M_{2^{p-} 3^{d+}}
+\frac{1}{4}\,\,\alpha_1^{d2+}\,\,M_{2^{p-} 3^{d+}}
+\frac{1}{4}\,\,\alpha_1^{p1-}\,\,M_{2^{p-} 2^{p-}}\,\, .
\hskip10.5ex 2^{p-}\\
\end{array}
\eqno (A29)$$

$\Sigma$ $\frac{7}{2} ^{+}$ $(8,4)$ $(70,2^+)$:

$$
\begin{array}{l}
\alpha_1^{s0+}=\lambda+
\frac{1}{4}\,\,\alpha_1^{s0s+}\,\, M_{1^{s+} 1^{s+}_s}
+\frac{1}{4}\,\,\alpha_1^{p2s+}\,\, M_{1^{s+} 3^{d+}_s}
+\frac{3}{4}\,\,\alpha_1^{d2s+}\,\, M_{1^{s+} 3^{d+}_s}
+\frac{3}{4}\,\,\alpha_1^{p1s-}\,\, M_{1^{s+} 2^{p-}_s}
\hskip4.0ex 1^{s+}\\
\\
\alpha_1^{p2+}=\lambda+
\frac{1}{4}\,\,\alpha_1^{s0s+}\,\, M_{3^{d+} 1^{s+}_s}
+\frac{1}{4}\,\,\alpha_1^{p2s+}\,\, M_{3^{d+} 3^{d+}_s}
+\frac{3}{4}\,\,\alpha_1^{d2s+}\,\, M_{3^{d+} 3^{d+}_s}
+\frac{3}{4}\,\,\alpha_1^{p1s-}\,\, M_{3^{d+} 2^{p-}_s}
\hskip4.0ex 3^{d+}\\
\\
\alpha_1^{s0s+}=\lambda+
\frac{1}{2}\,\,\alpha_1^{s0+}\,\, M_{1^{s+}_s 1^{s+}}
+\frac{1}{2}\,\,\alpha_1^{p2+}\,\, M_{1^{s+}_s 3^{d+}}
-\frac{1}{4}\,\,\alpha_1^{s0s+}\,\, M_{1^{s+}_s 1^{s+}_s}
-\frac{1}{4}\,\,\alpha_1^{p2s+}\,\,M_{1^{s+}_s 3^{d+}_s}
\hskip5.5ex 1^{s+}_s\\
\\
\hskip6ex
+\frac{3}{4}\,\,\alpha_1^{d2s+}\,\,M_{1^{s+}_s 3^{d+}_s}
+\frac{3}{4}\,\,\alpha_1^{p1s-}\,\,M_{1^{s+}_s 2^{p-}_s}\\
\\
\alpha_1^{p2s+}=\lambda+
\frac{1}{2}\,\,\alpha_1^{s0+}\,\, M_{3^{d+}_s 1^{s+}}
+\frac{1}{2}\,\,\alpha_1^{p2+}\,\, M_{3^{d+}_s 3^{d+}}
-\frac{1}{4}\,\,\alpha_1^{s0s+}\,\, M_{3^{d+}_s 1^{s+}_s}
-\frac{1}{4}\,\,\alpha_1^{p2s+}\,\,M_{3^{d+}_s 3^{d+}_s}
\hskip5.5ex 3^{d+}_s\\
\\
\hskip6ex
+\frac{3}{4}\,\,\alpha_1^{d2s+}\,\,M_{3^{d+}_s 3^{d+}_s}
+\frac{3}{4}\,\,\alpha_1^{p1s-}\,\,M_{3^{d+}_s 2^{p-}_s}\\
\\
\alpha_1^{d2s+}=\lambda+
\frac{1}{2}\,\,\alpha_1^{s0+}\,\, M_{3^{d+}_s 1^{s+}}
+\frac{1}{2}\,\,\alpha_1^{p2+}\,\, M_{3^{d+}_s 3^{d+}}
+\frac{1}{4}\,\,\alpha_1^{s0s+}\,\, M_{3^{d+}_s 1^{s+}_s}
+\frac{1}{4}\,\,\alpha_1^{p2s+}\,\,M_{3^{d+}_s 3^{d+}_s}
\hskip5.5ex 3^{d+}_s\\
\\
\hskip6ex
+\frac{1}{4}\,\,\alpha_1^{d2s+}\,\,M_{3^{d+}_s 3^{d+}_s}
+\frac{1}{4}\,\,\alpha_1^{p1s-}\,\,M_{3^{d+}_s 2^{p-}_s}\\
\\
\alpha_1^{p1s-}=\lambda+
\frac{1}{2}\,\,\alpha_1^{s0+}\,\, M_{2^{p-}_s 1^{s+}}
+\frac{1}{2}\,\,\alpha_1^{p2+}\,\, M_{2^{p-}_s 3^{d+}}
+\frac{1}{4}\,\,\alpha_1^{s0s+}\,\, M_{2^{p-}_s 1^{s+}_s}
+\frac{1}{4}\,\,\alpha_1^{p2s+}\,\,M_{2^{p-}_s 3^{d+}_s}
\hskip5.0ex 2^{p-}_s\\
\\
\hskip6ex
+\frac{1}{4}\,\,\alpha_1^{d2s+}\,\,M_{2^{p-}_s 3^{d+}_s}
+\frac{1}{4}\,\,\alpha_1^{p1s-}\,\,M_{2^{p-}_s 2^{p-}_s}\,\, .\\
\end{array}
\eqno (A30)$$

\newpage

$\Lambda$ $\frac{7}{2} ^{+}$ $(8,4)$ $(70,2^+)$:

$$
\begin{array}{l}
\alpha_1^{d2+}=\lambda+
\frac{1}{4}\,\,\alpha_1^{d2s+}\,\, M_{3^{d+} 3^{d+}_s}
+\frac{1}{4}\,\,\alpha_1^{p1s-}\,\, M_{3^{d+} 2^{p-}_s}
+\frac{3}{4}\,\,\alpha_1^{s0s+}\,\, M_{3^{d+} 1^{s+}_s}
+\frac{3}{4}\,\,\alpha_1^{p2s+}\,\, M_{3^{d+} 3^{d+}_s}
\hskip4.0ex 3^{d+}\\
\\
\alpha_1^{p1-}=\lambda+
\frac{1}{4}\,\,\alpha_1^{d2s+}\,\, M_{2^{p-} 3^{d+}_s}
+\frac{1}{4}\,\,\alpha_1^{p1s-}\,\, M_{2^{p-} 2^{p-}_s}
+\frac{3}{4}\,\,\alpha_1^{s0s+}\,\, M_{2^{p-} 1^{s+}_s}
+\frac{3}{4}\,\,\alpha_1^{p2s+}\,\, M_{2^{p-} 3^{d+}_s}
\hskip4.0ex 2^{p-}\\
\\
\alpha_1^{d2s+}=\lambda+
\frac{1}{2}\,\,\alpha_1^{d2+}\,\, M_{3^{d+}_s 3^{d+}}
+\frac{1}{2}\,\,\alpha_1^{p1-}\,\, M_{3^{d+}_s 2^{p-}}
+\frac{1}{2}\,\,\alpha_1^{s0s+}\,\, M_{3^{d+}_s 1^{s+}_s}
+\frac{1}{2}\,\,\alpha_1^{p2s+}\,\, M_{3^{d+}_s 3^{d+}_s}
\hskip5.5ex 3^{d+}_s\\
\\
\alpha_1^{p1s-}=\lambda+
\frac{1}{2}\,\,\alpha_1^{d2+}\,\, M_{2^{p-}_s 3^{d+}}
+\frac{1}{2}\,\,\alpha_1^{p1-}\,\, M_{2^{p-}_s 2^{p-}}
+\frac{1}{2}\,\,\alpha_1^{s0s+}\,\, M_{2^{p-}_s 1^{s+}_s}
+\frac{1}{2}\,\,\alpha_1^{p2s+}\,\, M_{2^{p-}_s 3^{d+}_s}
\hskip5.5ex 2^{p-}_s\\
\\
\alpha_1^{s0s+}=\lambda+
\frac{1}{2}\,\,\alpha_1^{d2+}\,\, M_{1^{s+}_s 3^{d+}}
+\frac{1}{2}\,\,\alpha_1^{p1-}\,\, M_{1^{s+}_s 2^{p-}}
+\frac{1}{6}\,\,\alpha_1^{d2s+}\,\, M_{1^{s+}_s 3^{d+}_s}
+\frac{1}{6}\,\,\alpha_1^{p1s-}\,\,M_{1^{s+}_s 2^{p-}_s}
\hskip6ex 1^{s+}_s\\
\\
\hskip6ex
+\frac{1}{3}\,\,\alpha_1^{s0s+}\,\,M_{1^{s+}_s 1^{s+}_s}
+\frac{1}{3}\,\,\alpha_1^{p2s+}\,\,M_{1^{s+}_s 3^{d+}_s}\\
\\
\alpha_1^{p2s+}=\lambda+
\frac{1}{2}\,\,\alpha_1^{d2+}\,\, M_{3^{d+}_s 3^{d+}}
+\frac{1}{2}\,\,\alpha_1^{p1-}\,\, M_{3^{d+}_s 2^{p-}}
+\frac{1}{6}\,\,\alpha_1^{d2s+}\,\, M_{3^{d+}_s 3^{d+}_s}
+\frac{1}{6}\,\,\alpha_1^{p1s-}\,\,M_{3^{d+}_s 2^{p-}_s}
\hskip5.5ex 3^{d+}_s\\
\\
\hskip6ex
+\frac{1}{3}\,\,\alpha_1^{s0s+}\,\,M_{3^{d+}_s 1^{s+}_s}
+\frac{1}{3}\,\,\alpha_1^{p2s+}\,\,M_{3^{d+}_s 3^{d+}_s}\,\, .\\
\end{array}
\eqno (A31)$$

\vskip2.0ex
{\bf Multiplet $(1,2)$.}
\vskip2.0ex

$\Lambda$ $\frac{5}{2} ^{+}$ $(1,2)$ $(70,2^+)$:

$$
\begin{array}{l}
\alpha_0^{s0+}=\lambda+
\frac{1}{4}\,\,\alpha_0^{s0s+}\,\, M_{0^{s+} 0^{s+}_s}
+\frac{1}{4}\,\,\alpha_0^{p2s+}\,\, M_{0^{s+} 2^{d+}_s}
+\frac{3}{4}\,\,\alpha_1^{d2s+}\,\, M_{0^{s+} 3^{d+}_s}
+\frac{3}{4}\,\,\alpha_1^{p1s-}\,\, M_{0^{s+} 2^{p-}_s}
\hskip4.0ex 0^{s+}\\
\\
\alpha_0^{p2+}=\lambda+
\frac{1}{4}\,\,\alpha_0^{s0s+}\,\, M_{2^{d+} 0^{s+}_s}
+\frac{1}{4}\,\,\alpha_0^{p2s+}\,\, M_{2^{d+} 2^{d+}_s}
+\frac{3}{4}\,\,\alpha_1^{d2s+}\,\, M_{2^{d+} 3^{d+}_s}
+\frac{3}{4}\,\,\alpha_1^{p1s-}\,\, M_{2^{d+} 2^{p-}_s}
\hskip4.0ex 2^{d+}\\
\\
\alpha_0^{s0s+}=\lambda+
\frac{1}{2}\,\,\alpha_0^{s0+}\,\, M_{0^{s+}_s 0^{s+}}
+\frac{1}{2}\,\,\alpha_0^{p2+}\,\, M_{0^{s+}_s 2^{d+}}
-\frac{1}{4}\,\,\alpha_0^{s0s+}\,\, M_{0^{s+}_s 0^{s+}_s}
-\frac{1}{4}\,\,\alpha_0^{p2s+}\,\,M_{0^{s+}_s 2^{d+}_s}
\hskip5.5ex 0^{s+}_s\\
\\
\hskip6ex
+\frac{3}{4}\,\,\alpha_1^{d2s+}\,\,M_{0^{s+}_s 3^{d+}_s}
+\frac{3}{4}\,\,\alpha_1^{p1s-}\,\,M_{0^{s+}_s 2^{p-}_s}\\
\\
\alpha_0^{p2s+}=\lambda+
\frac{1}{2}\,\,\alpha_0^{s0+}\,\, M_{2^{d+}_s 0^{s+}}
+\frac{1}{2}\,\,\alpha_0^{p2+}\,\, M_{2^{d+}_s 2^{d+}}
-\frac{1}{4}\,\,\alpha_0^{s0s+}\,\, M_{2^{d+}_s 0^{s+}_s}
-\frac{1}{4}\,\,\alpha_0^{p2s+}\,\,M_{2^{d+}_s 2^{d+}_s}
\hskip5.5ex 2^{d+}_s\\
\\
\hskip6ex
+\frac{3}{4}\,\,\alpha_1^{d2s+}\,\,M_{2^{d+}_s 3^{d+}_s}
+\frac{3}{4}\,\,\alpha_1^{p1s-}\,\,M_{2^{d+}_s 2^{p-}_s}\\
\\
\alpha_1^{d2s+}=\lambda+
\frac{1}{2}\,\,\alpha_0^{s0+}\,\, M_{3^{d+}_s 0^{s+}}
+\frac{1}{2}\,\,\alpha_0^{p2+}\,\, M_{3^{d+}_s 2^{d+}}
+\frac{1}{4}\,\,\alpha_0^{s0s+}\,\, M_{3^{d+}_s 0^{s+}_s}
+\frac{1}{4}\,\,\alpha_0^{p2s+}\,\,M_{3^{d+}_s 2^{d+}_s}
\hskip5.5ex 3^{d+}_s\\
\\
\hskip6ex
+\frac{1}{4}\,\,\alpha_1^{d2s+}\,\,M_{3^{d+}_s 3^{d+}_s}
+\frac{1}{4}\,\,\alpha_1^{p1s-}\,\,M_{3^{d+}_s 2^{p-}_s}\\
\\
\alpha_1^{p1s-}=\lambda+
\frac{1}{2}\,\,\alpha_0^{s0+}\,\, M_{2^{p-}_s 0^{s+}}
+\frac{1}{2}\,\,\alpha_0^{p2+}\,\, M_{2^{p-}_s 2^{d+}}
+\frac{1}{4}\,\,\alpha_0^{s0s+}\,\, M_{2^{p-}_s 0^{s+}_s}
+\frac{1}{4}\,\,\alpha_0^{p2s+}\,\,M_{2^{p-}_s 2^{d+}_s}
\hskip5.0ex 2^{p-}_s\\
\\
\hskip6ex
+\frac{1}{4}\,\,\alpha_1^{d2s+}\,\,M_{2^{p-}_s 3^{d+}_s}
+\frac{1}{4}\,\,\alpha_1^{p1s-}\,\,M_{2^{p-}_s 2^{p-}_s}\,\, .\\
\end{array}
\eqno (A32)$$

\newpage

{\large \bf References.}
\vskip5.0ex

\noindent
1. G.'t Hooft, Nucl. Phys. B{\bf 72} 461 (1974).

\noindent
2. E. Witten, Nucl. Phys. B{\bf 160} 57 (1979).

\noindent
3. J.L. Gervais and B. Sakita, Phys. Rev. Lett. {\bf 52} 87 (1984).

\noindent
4. R. Dashen and A.V. Manohar, Phys. Lett. B{\bf 315} 425 (1993).

\noindent
5. R. Dashen, E. Jenkins and A.V. Manohar, Phys. Rev. D{\bf 49} 4713 (1994).

\noindent
6. R. Dashen, E. Jenkins and A.V. Manohar, Phys. Rev. D{\bf 51} 3697 (1995).

\noindent
7. C.D. Carone, H. Georgi and S. Osofsky, Phys. Lett. B{\bf 322} 227 (1994).

\noindent
8. M.A. Luty and J. March-Russell, Nucl. Phys. B{\bf 426} 71 (1994).

\noindent
9. E. Jenkins, Phys. Lett. B{\bf 315} 441 (1993).

\noindent
10. E. Jenkins and R.F. Lebed, Phys. Rev. D{\bf 52} 282 (1995).

\noindent
11. J. Dai, R. Dashen, E. Jenkins, and A. V. Manohar,
Phys. Rev. D{\bf 53} 273 (1996).

\noindent
12. N. Matagne and Fl. Stancu, Phys. Rev. D{\bf 71} 015710 (2005).

\noindent
13. N. Matagne and Fl. Stancu, arXiv: hep-ph/0610099.

\noindent
14. N. Matagne and Fl. Stancu, Phys. Lett. B{\bf 631} 7 (2005).

\noindent
15. N. Matagne and Fl. Stancu, Phys. Rev. D{\bf 74} 034014 (2006).

\noindent
16. N. Matagne and Fl. Stancu, arXiv: hep-ph/0603092.

\noindent
17. I.J.R. Aitchison, J. Phys. G{\bf 3} 121 (1977).

\noindent
18. J.J. Brehm, Ann. Phys. (N.Y.) {\bf 108} 454 (1977).

\noindent
19. I.J.R. Aitchison and J.J. Brehm, Phys. Rev. D{\bf 17} 3072 (1978).

\noindent
20. I.J.R. Aitchison and J.J. Brehm, Phys. Rev. D{\bf 20} 1119, 1131 (1979).

\noindent
21. J.J. Brehm, Phys. Rev. D{\bf 21} 718 (1980).

\noindent
22. S.M. Gerasyuta, Yad. Fiz. {\bf 55} 3030 (1992).

\noindent
23. S.M. Gerasyuta, Z. Phys. C{\bf 60} 683 (1993).

\noindent
24. S.M. Gerasyuta, E.E. Matskevich, Yad. Fiz. {\bf 11} (2007),
arXiv: hep-ph/0701120.

\noindent
25. W.M. Yao et al. (Particle Data Group), J. Phys. G {\bf 33}, 1 (2006).

\noindent
26. A.De Rujula, H.Georgi and S.L.Glashow, Phys. Rev. D{\bf 12} 147 (1975).

\noindent
27. V.V. Anisovich, S.M. Gerasyuta and A.V. Sarantsev,
Int. J. Mod. Phys. A{\bf 6} 625 (1991).

\noindent
28. G. Chew and S. Mandelstam, Phys. Rev. {\bf 119} 467 (1960).

\noindent
29. V.V. Anisovich and A.A. Anselm, Usp. Fiz. Nauk {\bf 88} 287 (1966).

\noindent
30. S.M. Gerasyuta and I.V. Keltuyala, Yad. Fiz. {\bf 54} 793 (1991).

\noindent
31. S.M. Gerasyuta and I.V. Kochkin, Int. J. Mod. Phys. E{\bf 15} 71 (2006).

\noindent
32. S.M. Gerasyuta and I.V. Kochkin, Phys. Rev. D{\bf 66} 116001 (2002).

\noindent
33. S.M. Gerasyuta and D.V. Ivanov, Vest. SPb University
Ser. 4 {\bf 11} 3 (1996).

\noindent
34. N. Isgur and G. Karl, Phys. Rev. D{\bf 19} 2653 (1979).

\noindent
35. A.J. Hey and R.L. Kelly, Phys. Rept. {\bf 96} 71 (1983).

\noindent
36. S. Capstick and N. Isgur, Phys. Rev. D{\bf 34} 2809 (1986).

\noindent
37. S. Capstick and W. Roberts, nucl-th/000828.

\end{document}